\documentclass[apj,numberedappendix,twocolappendix,appendixfloats]{emulateapj}

\usepackage{rotating}
\usepackage{epsfig}

\usepackage[fleqn]{amsmath}

\newcommand{\lt}{\ifmmode\,<\,\else \,$<$\,\fi}
\newcommand{\kms}{\ifmmode\,{\rm km}\,{\rm s}^{-1}\else km$\,$s$^{-1}$\fi}

\newcommand{\magarc}{\ifmmode {{{{\rm mag}~{\rm arcsec}}^{-2}}}
             \else {{{mag}$~${arcsec}$^{-2}$}}
             \fi}

\newcommand{\oh}{$12+\log{\mathrm{O/H}}$}

\newcommand{\funit}{\mathrm{ergs}~\mathrm{s}^{-1}~\mathrm{cm}^{-2}}
\newcommand{\Msun}{\mathrm{M}_{\sun}}

\newcommand{\Hii}{H~{\sc ii}}
\newcommand{\Oii}{[O~{\sc ii}]}
\newcommand{\Oiii}{[O~{\sc iii}]}
\newcommand{\Neiii}{[Ne~{\sc iii}]}
\newcommand{\Ha}{H$\alpha$}
\newcommand{\Hb}{H$\beta$}
\newcommand{\Hg}{H$\gamma$}
\newcommand{\Hd}{H$\delta$}
\newcommand{\Nii}{[N~{\sc ii}]}
\newcommand{\Sii}{[S~{\sc ii}]}
\newcommand{\Siii}{[S~{\sc iii}]}
\newcommand{\ROiii}{R$_{\mathrm{ [O~\textsc{iii}]}}$}

\newcommand{\Te}{T$_{\mathrm e}$}
\newcommand{\Ne}{n$_{\mathrm e}$}

\usepackage{bbm}

\shorttitle{Direct Metallicity Measurements at $z=0.8$}

\begin{document}

\title{Temperature-based metallicity measurements at $z=0.8$: \\
direct calibration of strong-line diagnostics at intermediate redshift}


\author{Tucker~Jones\altaffilmark{1}}
\author{Crystal~Martin\altaffilmark{1}}
\author{Michael~C.~Cooper\altaffilmark{2}}

\altaffiltext{1} {Department of Physics, Broida Hall, University of California, Santa Barbara, CA 93106, USA}
\altaffiltext{2} {Center for Cosmology, Department of Physics and Astronomy, 4129 Reines Hall, University of California, Irvine, CA 92697, USA}


\begin{abstract}

We present the first direct calibration of strong-line metallicity diagnostics at significant cosmological distances using a sample at $z\simeq0.8$ drawn from the DEEP2 Galaxy Redshift Survey. Oxygen and neon abundances are derived from measurements of electron temperature and density. We directly compare various commonly used relations between gas-phase metallicity and strong line ratios of O, Ne, and H at $z\simeq0.8$ and $z=0$. There is no evolution with redshift at high precision ($\Delta \log{\mathrm{O/H}} = -0.01\pm0.03$, $\Delta \log{\mathrm{Ne/O}} = 0.01 \pm 0.01$). O, Ne, and H line ratios follow the same locus at $z\simeq0.8$ as at $z=0$ with $\lesssim$0.02 dex evolution and low scatter ($\lesssim$0.04 dex). This suggests little or no evolution in physical conditions of \Hii\ regions at fixed oxygen abundance, in contrast to models which invoke more extreme properties at high redshifts.
We speculate that offsets observed in the \Nii/Ha\ versus \Oiii/\Hb\ diagram at high redshift are therefore due to \Nii\ emission, likely as a result of relatively high N/O abundance. If this is indeed the case, then nitrogen-based metallicity diagnostics suffer from systematic errors at high redshift. Our findings indicate that locally calibrated abundance diagnostics based on $\alpha$-capture elements can be reliably applied at $z\simeq1$ and possibly at much higher redshifts.
This constitutes the first firm basis for the widespread use of empirical calibrations in high redshift metallicity studies.

\end{abstract}


\section{Introduction}\label{sec:intro}

Gas-phase oxygen abundance (hereafter ``metallicity") is a valuable yet elusive diagnostic of galaxy evolution. Metallicity is driven by the production of heavy elements via star formation, and modulated by gaseous inflows and outflows \citep[e.g.,][]{Tinsley1978,Edmunds1990,Erb2008,Finlator2008,Peeples2011}. Precise measurements of metallicity and its evolution with time, especially in combination with accumulated stellar mass and gas content, therefore provide information on the history of gas flows which regulate galaxy mass growth. This prospect has motivated vast efforts to characterize galaxy metallicity as a function of stellar mass and cosmic time \citep[][{\em and many others}]{Erb2006,Maiolino2008,Mannucci2010,Richard2011,Belli2013,Henry2013,Zahid2013,Cullen2014,Wuyts2014,Sanders2015}, and to interpret the resultant data using theoretical models incorporating cosmologically-motivated accretion and feedback \citep[e.g.,][]{Brooks2007,Finlator2008,Dave2011,Brook2012,Velona2013,Obreja2014}. Observationally there exists a mass-metallicity relation in the sense that galaxies with lower stellar masses have lower metallicity  on average \citep[e.g.,][]{Tremonti2004, Lequeux1979}. This is commonly attributed to metal-enriched outflows which are more efficient at removing gas from low-mass systems. The mass-metallicity relation evolves toward lower metallicity at higher redshifts \citep[e.g.,][]{Maiolino2008,Zahid2013}, roughly commensurate with the increased gas fractions measured for modest samples \citep{Tacconi2013}.

Although progress towards quantifying the chemical evolution of galaxies over cosmic time is seemingly impressive, essentially all of the data-driven results described above rely on indirect estimates of metallicity and should be regarded with due skepticism. True measurements of nebular ionic abundances require observations of either recombination lines, or collisionally excited emission lines combined with knowledge of the temperature and density \citep[e.g.,][]{Aller1959,Tsamis2003,Stasinska2004}, although we caution that these methods are still sensitive to assumptions such as the electron energy distribution \citep[e.g.,][]{Nicholls2012}.
This presents an observational challenge in that temperature-sensitive and recombination lines of heavy elements are fainter than the Balmer recombination lines of hydrogen by factors of $\gtrsim$100 and $\gtrsim$1000, respectively. This has motivated the development of more practical -- although more uncertain -- methods based on the flux ratios of Balmer lines and comparably bright collisionally excited metal lines. These ``strong line" methods, first introduced by \cite{Jensen1976} and \cite{Pagel1979}, exploit the fact that various flux ratios are strongly correlated with direct measurements of metallicity. Relations between strong line ratios and metallicity can be calibrated either empirically from direct measurements of bright nearby objects, or through detailed photoionization modeling \citep[e.g.,][]{Kewley2002,Dopita2013}. Various such calibrations are now in widespread use and are virtually the {\em only} method used to estimate galaxy metallicities at high redshift.

Strong line methods can in principle be used to infer metallicities with an accuracy of $\sigma(\log{\mathrm{O/H}})\sim0.1$--0.2 dex \citep[although uncertainty varies depending on the line ratio and ionization regime; e.g.,][]{Pettini2004,Maiolino2008}. This uncertainty floor is limited by intrinsic scatter in metallicity at fixed strong line ratios, which arises from variations in physical properties such as the ionization parameter.
In practice, however, there is no clear consensus on the absolute metallicity scale: different published calibrations differ by up to 0.7 dex \citep{Kewley2008}! Nonetheless these discrepancies have not curtailed efforts to quantify {\em relative} metallicity evolution of galaxies up to redshifts $z>3$ using consistent sets of calibrations. Another concern is whether strong line methods can be reliably applied at high redshifts at all, given that existing calibrations are based entirely on local galaxies and \Hii\ regions. If the physical state of \Hii\ regions evolves systematically with redshift, this will induce artificial evolutionary trends in metallicities inferred from locally-calibrated diagnostics \citep[e.g.,][]{Kewley2013}.

Large spectroscopic surveys have now conclusively shown that emission line ratios of high redshift galaxies ($z\gtrsim1$) are significantly offset from the locus formed by local galaxies. This is at least true for the ``BPT diagram" (named for the authors \citealt{Baldwin1981}) of \Oiii/\Hb\ versus \Nii/\Ha\ for which the typical offset at $z\simeq2.3$ is 0.2--0.4 dex \citep{Shapley2015,Steidel2014}. Other diagrams involving only oxygen, hydrogen, and sulfur lines do not show significant offsets, however. 
Since \Nii/\Ha\ and O3N2~$=$~(\Oiii/\Hb)/(\Nii/Ha) are among the most commonly used strong-line metallicity indicators at high redshift \citep[e.g.,][]{Erb2006,Steidel2014,Sanders2015}, their intrinsic evolution necessitates a quantitative revision of many previous results. Several possible causes for this evolution have been proposed \citep[e.g.,][]{Kewley2013} and there is some evidence that elevated electron density may be at least partially responsible \citep{Hainline2009,Shirazi2014}. 
Increased stellar effective temperatures can also explain the evolution \citep[as considered in detail by][]{Steidel2014}, although this should result in similar offsets in \Sii/\Ha\ versus \Oiii/\Hb, which are not observed \citep{Shapley2015}.
In some cases the offsets are explained by a combination of star formation and AGN excitation \citep{Wright2010,Newman2014}. However, integral field spectroscopic observations of lensed galaxies show that these offsets are present even for individual giant \Hii\ regions \citep{Jones2010,Jones2013}. At the very least it is clear that metallicities inferred from these diagnostics are not consistent with local samples nor even internally among high redshift samples \citep[e.g.,][]{Steidel2014}.

Our study is motivated primarily by measurements of strong evolution in the BPT line ratio diagram and the associated unknown astrophysics. Ultimately we wish to understand the physical origin of evolution in nebular emission spectra, and to determine which (if any) strong line metallicity diagnostics can be calibrated for reliable use at high redshifts. These goals require accurate measurements of the physical properties of high redshift galaxies, particularly metallicity. In this paper we analyze a sample of galaxies at $z\simeq0.8$ for which we obtain sensitive measurements of electron density and temperature from the nebular emission spectra. This provides direct metallicities which we compare with local galaxies analyzed in exactly the same manner. Our analysis relies principally on the temperature-sensitive \Oiii$\lambda$4363/\Oiii$\lambda\lambda$4959,5007 flux ratio. Measurements of the weak \Oiii$\lambda$4363 line have previously been reported for modest samples of intermediate-redshift galaxies \citep{Hoyos2005,Kakazu2007,Ly2014a,Ly2014,Amorin2014} as well as a handful at $z>1$ \citep{Yuan2009,Christensen2012, Brammer2012b,Stark2013,James2014,Maseda2014}, but no effort has yet been attempted to calibrate strong line metallicity diagnostics beyond the local universe. This paper represents a significant first step and we hope that our work can soon be extended with well-defined samples at higher redshifts.

This paper is structured as follows. Section~\ref{sec:sample} presents our sample selection and a brief discussion of sample bias mitigation. Measurements of physical properties (nebular reddening, electron temperature, electron density, and metallicity) are described in Section~\ref{sec:properties}. Strong line metallicity calibrations are calculated in Section~\ref{sec:diagnostics}. Section~\ref{sec:results} quantitatively compares the (lack of) redshift evolution in relations between observed emission line ratios and derived physical properties; this section constitutes the primary results of our work. We discuss some interesting implications and briefly summarize our findings in Sections~\ref{sec:discussion} and \ref{sec:conclusions} respectively. Throughout the paper we make use of the following notation conventions: unless stated otherwise, \Oii, \Oiii, and \Neiii\ refer to \Oii$\lambda\lambda$3727,3729, \Oiii$\lambda$5007, and \Neiii$\lambda$3869 respectively. Expressions such as \Oiii/\Oii\ refer to the emission line flux ratios. All magnitudes are in the AB system \citep{Oke1974}.

\section{Sample selection and analysis}\label{sec:sample}

We have carefully selected a sample of distant galaxies from the DEEP2 Galaxy Redshift Survey \citep[DEEP2;][]{Davis2003,Newman2013} which are suitable for temperature-based metallicity analysis. DEEP2 obtained redshifts for $>$50,000 galaxies at up to $z\lesssim1.4$ using the DEIMOS spectrograph on the Keck II telescope. The spectra typically span a wavelength range of $\sim$6500--9100 \AA\ with resolution R $\simeq 5000$. All spectra are corrected for telluric absorption and system throughput using methods developed by the DEEP2 team \citep{Cooper2012}, and corrected for Milky Way extinction using the dust map of \cite{Schlegel1998}. We examined all galaxy spectra with secure redshifts (DEEP2 quality code Q~$=3$ or 4) such that the wavelength coverage spans at least 3626--4960~\AA\ in the rest frame, thereby including strong emission lines of \Oii\ and \Oiii. For initial selection, we fit single Gaussian profiles to \Oiii$\lambda\lambda$4959,5007 and require a combined signal-to-noise $>$80 (or $>$20 in \Oiii$\lambda$4959 for cases where the 5007 line is not covered). This yields 254 spectra with redshifts $0.72<z<0.87$. After removing duplicate observations, spectra where a crucial emission line (either \Hg\ or \Oiii$\lambda$4363) fell in the DEIMOS chip gap, and spectra strongly affected by telluric absorption residuals, this sample is reduced to 196 galaxies. We identify 10 of these as containing an active galactic nucleus (AGN) on the basis of broad emission lines.

For each spectrum we measure the flux and uncertainty of various emission lines using Gaussian profile fits. The lines of interest for this paper are \Oiii$\lambda\lambda$4959,5007, \Oiii$\lambda$4363, \Neiii$\lambda$3869, \Oii$\lambda\lambda$3727,3729, \Hb, \Hg, and \Hd. All lines are fit with a single component except for the \Oii\ doublet, which we fit with a double Gaussian function using the total flux and doublet ratio as free parameters. In all cases we fix the redshift and line width to the values derived at high signal-to-noise from \Oiii$\lambda\lambda$4959,5007. This maximizes the precision of flux measurements for fainter lines. In the $\sim$30\% of cases where \Oiii$\lambda$5007 is not covered, we adopt a flux ratio \Oiii$\lambda$5007/\Oiii$\lambda$4959 $=3$.

The hydrogen Balmer lines are affected by underlying stellar absorption, such that emission fluxes measured from a single-component fit will underestimate the true value. In general the stellar continuum is not strong enough to robustly measure Balmer absorption profiles for individual galaxies. We therefore constructed a composite from the median of all spectra, normalized to the continuum flux at rest frame 4150--4300 \AA. We fit the composite spectrum with a linear combination of simple stellar population templates from the library of \cite{Gonzalez1999}, using the penalized pixel fitting method of \cite{Cappellari2004}. The resulting average stellar spectrum is scaled to the continuum level of each galaxy in our sample and subtracted, and we re-fit the Balmer lines to determine the level of stellar absorption. The median correction for stellar absorption is 1.1--1.2 \AA\ in rest frame equivalent width for \Hb, \Hg, and \Hd. We correct the flux of each object for this absorption and include a conservative uncertainty of $\pm0.6$ \AA\ (rest frame) to account for sample variance. This additional uncertainty is propagated throughout the analysis. The stellar absorption correction is typically small (median $\sim$6\% in \Hd\ and $\sim$2\% in \Hb) due to the large Balmer emission equivalent widths.

\subsection{The \Te\ sample}

\begin{figure}
\includegraphics[width=\columnwidth]{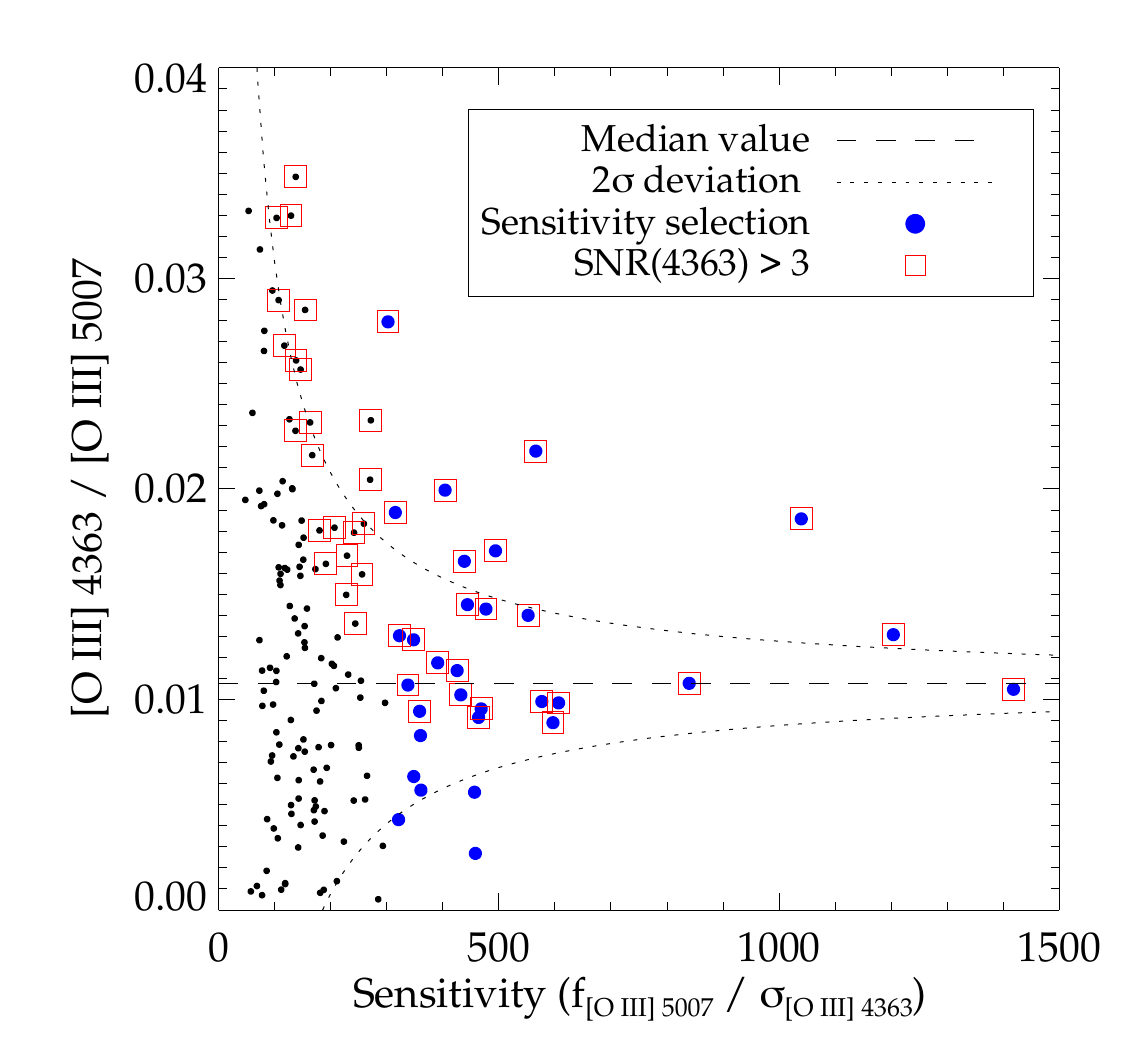}
\caption{
\label{fig:sample}
Sample selection and bias mitigation. Small black points show the parent sample of 186 star forming galaxies drawn from the DEEP2 survey. The data span a wide range of temperature-sensitive \Oiii\ flux ratios. The sample median and $2\sigma$ statistical deviation are shown by dashed and dotted lines, respectively, demonstrating that scatter in \Oiii\ ratios is largely (though not entirely) due to noise. To reduce the effects of random noise, we select a sub-sample above a limiting sensitivity in the \Oiii\ line ratio, shown by the larger blue points. We also show the sub-sample for which \Oiii$\lambda$4363 is detected at $\geq3\sigma$ significance (open squares). Selecting based on \Oiii$\lambda$4363 significance would clearly induce a strong bias toward higher \ROiii\ (and therefore higher temperature and lower metallicity). Our sensitivity-based selection eliminates this bias and ensures a representative distribution of derived metallicities.
}
\end{figure}

\begin{figure}
\includegraphics[width=\columnwidth]{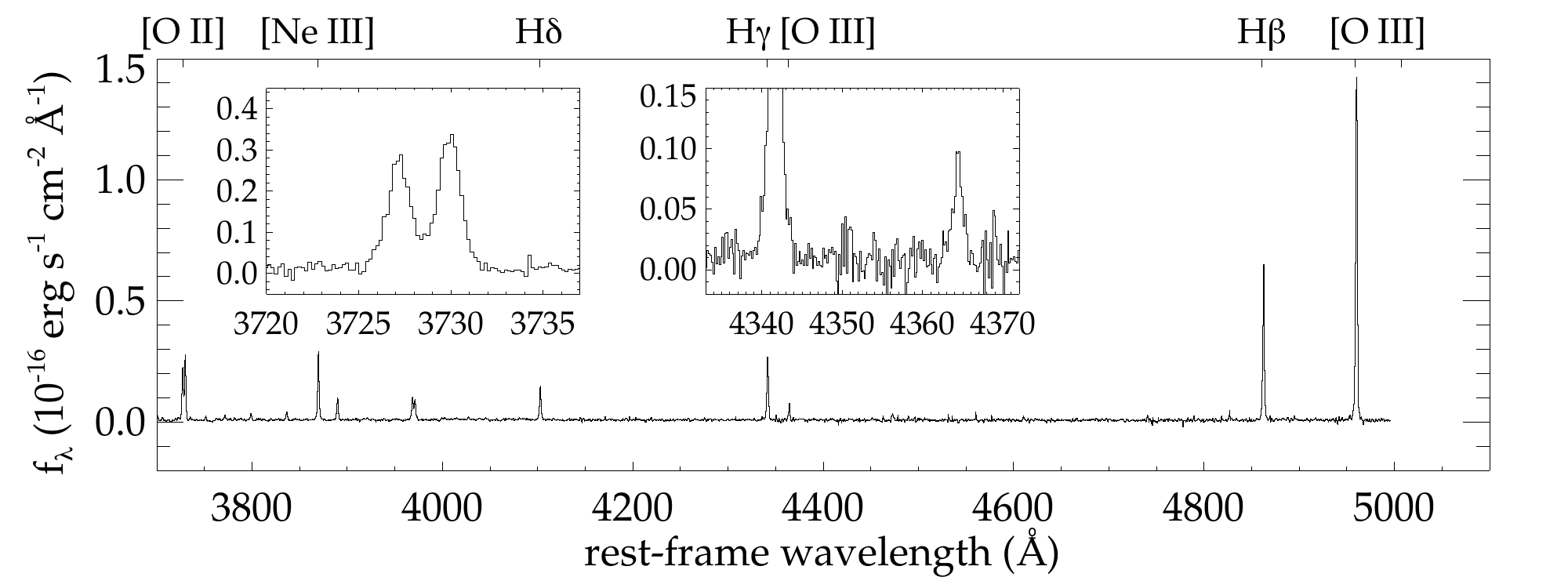}
\includegraphics[width=\columnwidth]{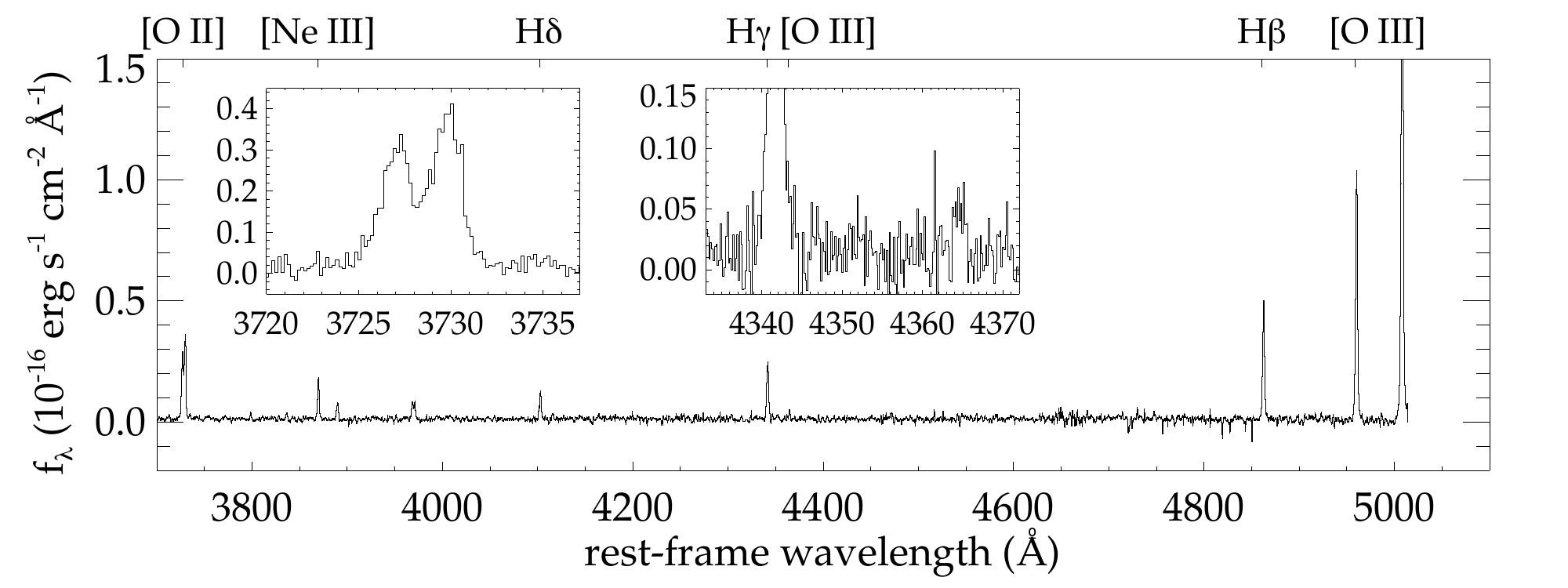}
\includegraphics[width=\columnwidth]{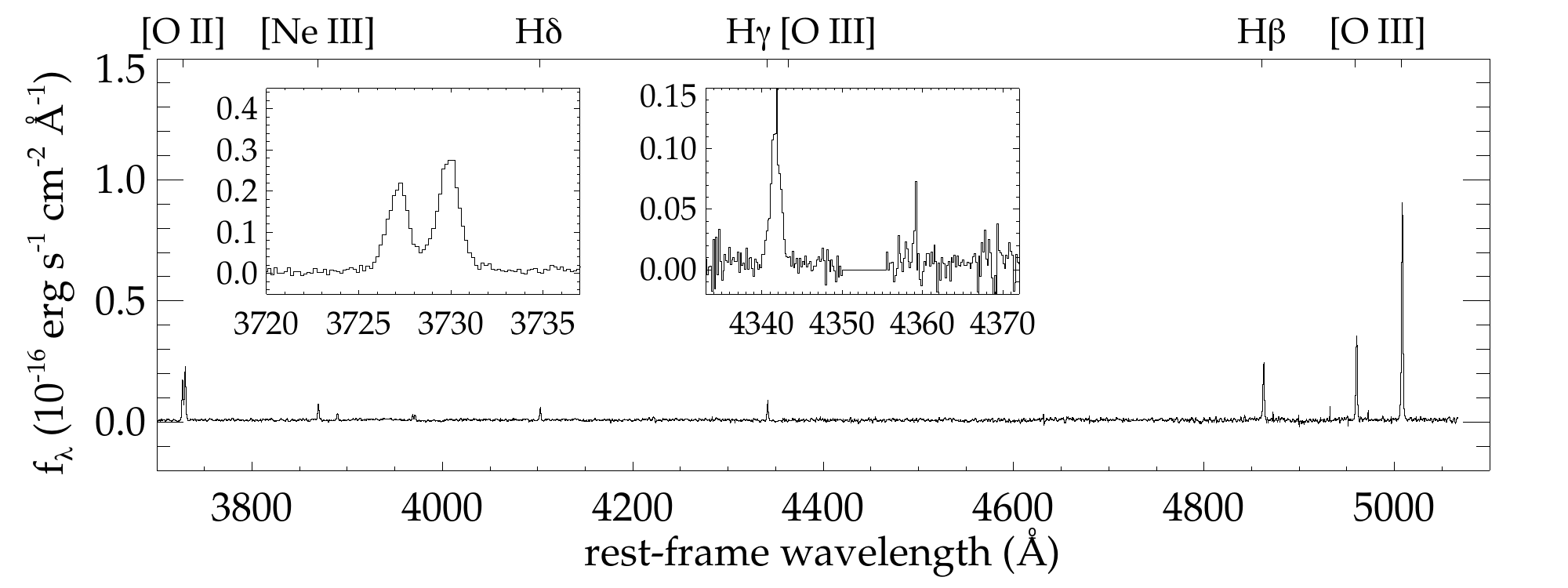}
\caption{
\label{fig:examples}
Spectra from the \Te\ sample with the highest (top), median (middle), and lowest (bottom) formal signal-to-noise ratio in \Oiii$\lambda$4363. These examples are chosen to show the range of data quality. Various salient characteristics are apparent such as the large Balmer emission equivalent widths, high \Oiii/\Hb\ flux ratios, and well-resolved \Oii$\lambda\lambda$3727,3729 doublet ratios indicative of low electron densities ($\lesssim 200$ cm$^{-3}$).
}
\end{figure}

The goals of this work require a uniformly selected sample with accurate measurements of electron temperature and density from nebular emission line ratios. The parent DEEP2 sample of 186 star forming galaxies spans a range of emission line strengths, with more than an order of magnitude variation in the detection significance of \Oiii$\lambda\lambda$4959,5007. Consequently there is a wide range in measurement precision for the temperature-sensitive ratio \ROiii\ $=$ (\Oiii$\lambda$4363)$/$(\Oiii$\lambda\lambda$4959,5007), illustrated in Figure~\ref{fig:sample}. We find a median flux ratio \Oiii$\lambda$5007/\Oiii$\lambda$4363~$\simeq 100$ (Figure~\ref{fig:sample}). 
In the remainder of this paper we therefore predominantly analyze the sub-sample for which \Oiii$\lambda$5007 flux is $\geq$300 times larger than the uncertainty in \Oiii$\lambda$4363 flux (i.e., the ratio \ROiii\ is measured with 1$\sigma$ precision $\le$0.0025), which we refer to as the ``\Te\ sample."
This ensures a $>$3$\sigma$ detection of \Oiii$\lambda$4363 in most cases, although the actual significance is subject to both noise fluctuations and intrinsic physical variation. Ultimately this selection yields a sample of 32 galaxies with a median \Oiii$\lambda$4363 significance of 5.3$\sigma$. Example spectra spanning the range of \Oiii$\lambda$4363 fluxes are shown in Figure~\ref{fig:examples}. The following analysis is based on this sample of 32 galaxies unless stated otherwise.

\begin{figure}
\includegraphics[width=\columnwidth]{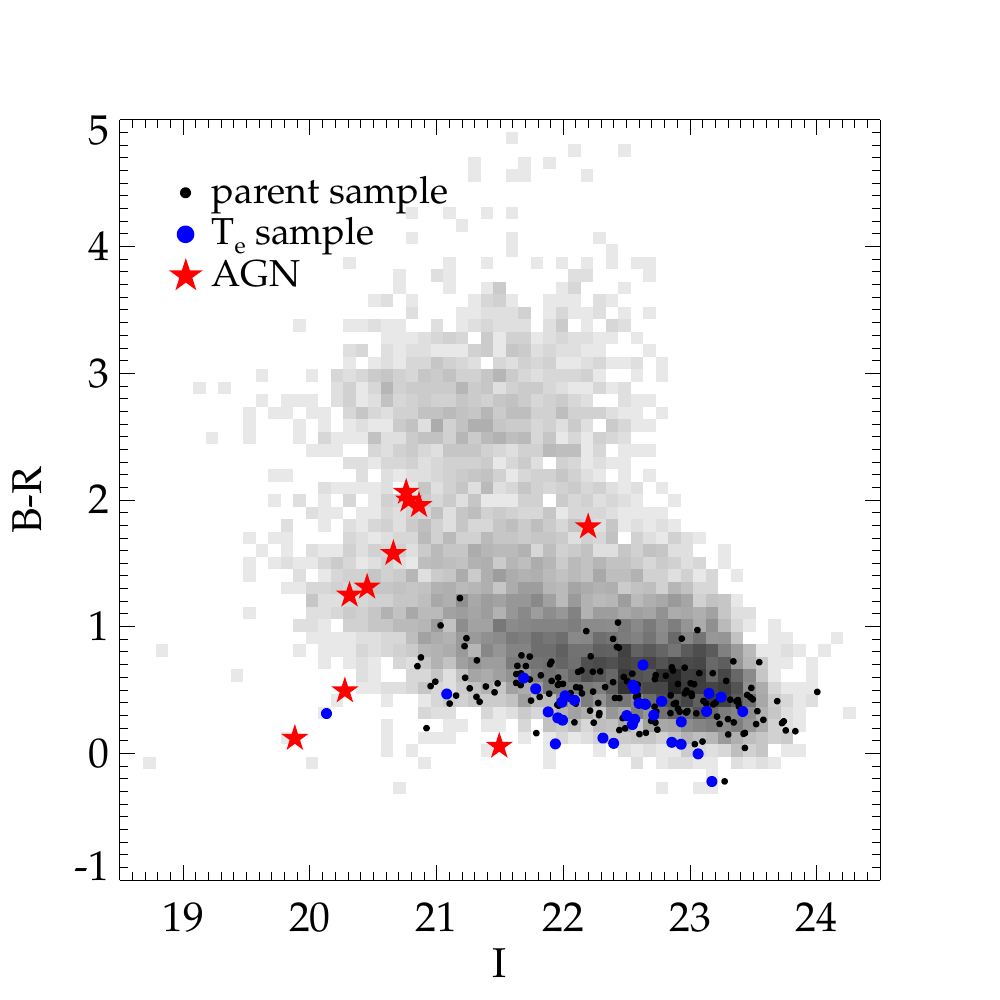}
\caption{
\label{fig:cmd}
Color-magnitude diagram of galaxies with secure $0.7<z<0.9$ from the DEEP2 survey, using the photometric measurements presented in \cite{Coil2004}. B, R, and I magnitudes refer to the observed passbands (and apparent color) rather than the rest frame.
Grey shading represents the square root of the number of objects in each bin. A clear bimodality is apparent, with the passive ``red sequence" at B--R~$\gtrsim2$ and the star-forming ``blue cloud" at B--R~$\lesssim2$. Galaxies studied in this paper populate the blue cloud, with blue colors and approximately representative luminosities relative to the overall star forming population.
}
\end{figure}

To place our sample in the context of the general galaxy population, we show a color-magnitude diagram in Figure~\ref{fig:cmd}. The \Te\ sample lies in the star-forming ``blue cloud" with representative luminosities and relatively blue colors compared to the population of star-forming galaxies at $z\simeq0.8$. We refer interested readers to \cite{Ly2014} for a more detailed discussion of stellar population properties and their relation to metallicity, which is beyond the scope of this paper. The \Te\ sample is similar to the 28 DEEP2 galaxies analyzed by \cite{Ly2014}, and indeed our samples overlap by $\sim$50\%. The differences are that \cite{Ly2014} require a $>$3$\sigma$ detection of \Oiii$\lambda$4363 and coverage of \Oiii$\lambda$5007, and do not impose a restriction on \ROiii\ precision. In terms of the \cite{Ly2014} sample, those galaxies which also appear in the \Te\ sample have representative stellar masses ($\simeq$10$^8 - 10^9~\Msun$) and relatively high metallicities, reflecting the different selection criteria.

\subsection{Local comparison sample}

We require an appropriate $z=0$ reference sample to analyze the degree of evolution in physical properties. For this purpose we use the measurements of \cite{Izotov2006}, whose galaxy sample is selected from Data Release 3 of the Sloan Digital Sky Survey \citep[SDSS;][]{Abazajian2005} based on \Hb\ flux ($\geq10^{14}~\funit$) and lack of nuclear activity. 
While the selection is not identical to the DEEP2 sample, it is similarly based on strong emission line fluxes and, crucially, does not explicitly select for detection of \Oiii$\lambda$4363. We analyze only the 113 spectra (at median $z=0.05$) with coverage of \Oii$\lambda\lambda$3727,3729 and with \Oiii$\lambda$5007 fluxes $\geq$300 times larger than the uncertainty in \Oiii$\lambda$4363 flux. This ensures good precision and consistency with the $z\simeq0.8$ sample, although the full \cite{Izotov2006} sample gives consistent results.

For purposes of comparing strong emission line ratios, we also utilize the MPA-JHU catalog\footnote{http://www.mpa-garching.mpg.de/SDSS/DR7/} of spectral measurements from Data Release 7 of the SDSS \citep{Abazajian2009}. We identify star forming galaxies according to their \Nii/\Ha\ and \Oiii/\Hb\ flux ratios using the classification scheme of \cite{Kauffmann2003} and select a sub-sample for which the following relevant lines are all detected at $\geq$10$\sigma$ significance: \Oii$\lambda\lambda$3727,3729, \Neiii$\lambda$3869, \Hd, \Hg, \Hb, \Ha, \Nii$\lambda$6584, and \Oiii$\lambda$5007. This yields $\sim$20,000 spectra. Low-level AGN activity is increasingly difficult to diagnose for flux ratios below \Oiii/\Hb~$\lesssim1$, however the spectra analyzed in this work have \Oiii/\Hb~$>2$ in all cases. The SDSS sample shown e.g. in Figure~\ref{fig:o3bias} is therefore representative of star forming galaxies in the local volume for the relevant range of emission line ratios.

\subsection{Sample bias}

\begin{figure}
\includegraphics[width=\columnwidth]{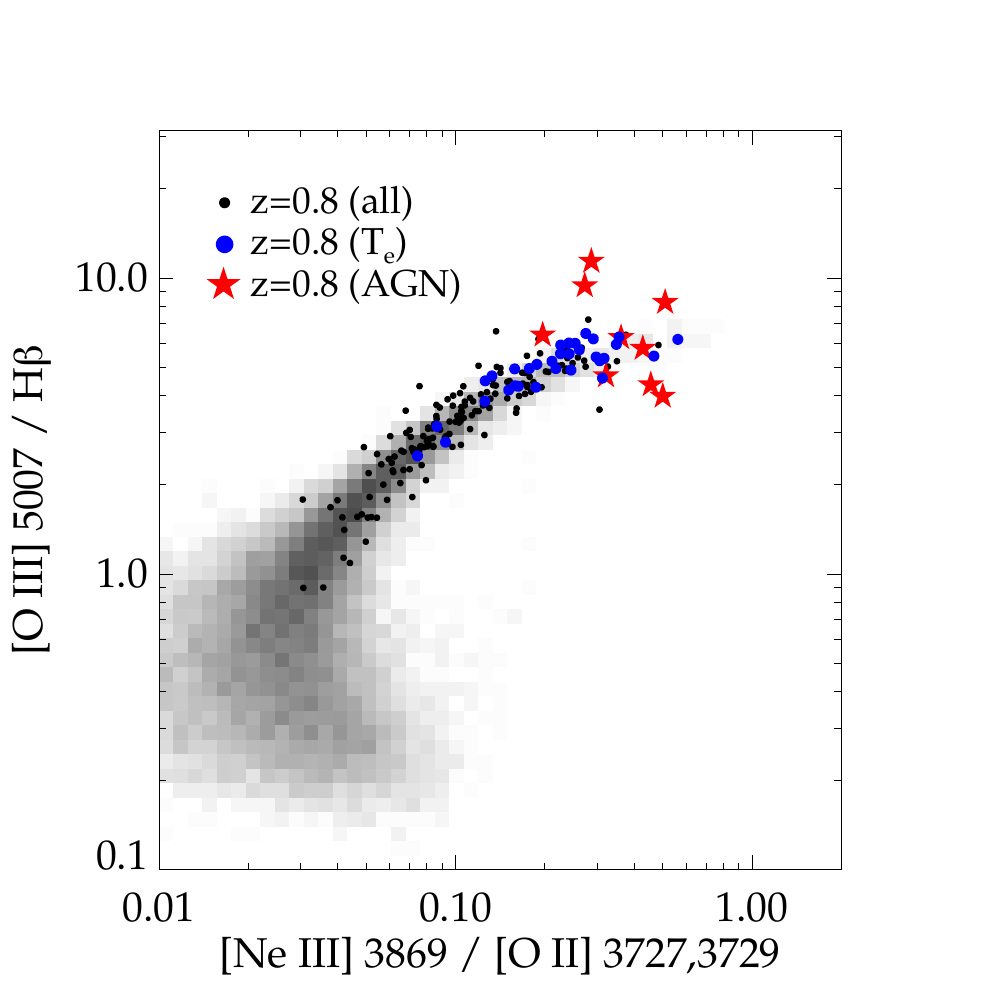}
\caption{
\label{fig:o3bias}
Strong emission line ratios at $z\simeq0.8$ compared to local galaxies. Grey shading shows the population of SDSS galaxies which are classified as star forming according to \cite{Kauffmann2003}. The increased scatter and general trend at \Oiii/\Hb~$\lesssim1$ may be due in part to a mix of star formation and AGN excitation in these galaxies; the ionizing source is increasingly ambiguous at lower \Oiii/\Hb. Our $z\simeq0.8$ sample is consistent with the local star-forming locus, lying exclusively at relatively high ratios of \Oiii/\Hb\ and \Neiii/\Oii. The \Te\ subsample has even higher line ratios on average but likewise shows no offset from the locus of galaxies at $z\simeq0$, nor from the broader population at $z\simeq0.8$. Broad-line AGN in the parent $z\simeq0.8$ sample lie near the locus of star formation with a considerably larger scatter.
}
\end{figure}

We have been careful to construct a sample without imposing any explicit restriction on \Oiii$\lambda$4363 flux. Figure~\ref{fig:sample} illustrates that requiring a detection of \Oiii$\lambda$4363 would induce a bias toward artificially high \ROiii\, and hence higher derived temperatures and lower metallicities. Selecting on the basis of strong-line (\Oiii$\lambda\lambda$4959,5007) flux and spectral sensitivity simultaneously mitigates this bias and ensures good constraints on \Te\ and metallicity, even for cases with very weak \Oiii$\lambda$4363 emission. To check whether the adopted sample selection may introduce other biases, we plot strong line ratios of \Oiii/\Hb\ versus \Neiii/\Oii\ in Figure~\ref{fig:o3bias}. The \Te\ subsample has higher average \Oiii/\Hb\ compared to the parent sample, as one might expect given the selection based on \Oiii\ flux. However there appears to be no systematic bias in the sense that the \Te\ subsample has \Oiii/\Hb\ ratios which are normal considering their other properties (e.g., \Neiii/\Oii\ and other emission line ratios). Likewise the local comparison sample is selected on the basis of strong emission lines without regard to \Oiii$\lambda$4363 \citep{Izotov2006}, and the line ratios indicate no significant systematic offset compared to the larger SDSS sample. It is also apparent in Figure~\ref{fig:o3bias} that star forming galaxies at $z\simeq0.8$ lie along the same locus as local galaxies in terms of these line ratios, with no significant offset and comparable scatter.

\section{Physical properties}\label{sec:properties}

\subsection{Nebular extinction and reddening}\label{sec:extinction}

\begin{figure}
\includegraphics[width=\columnwidth]{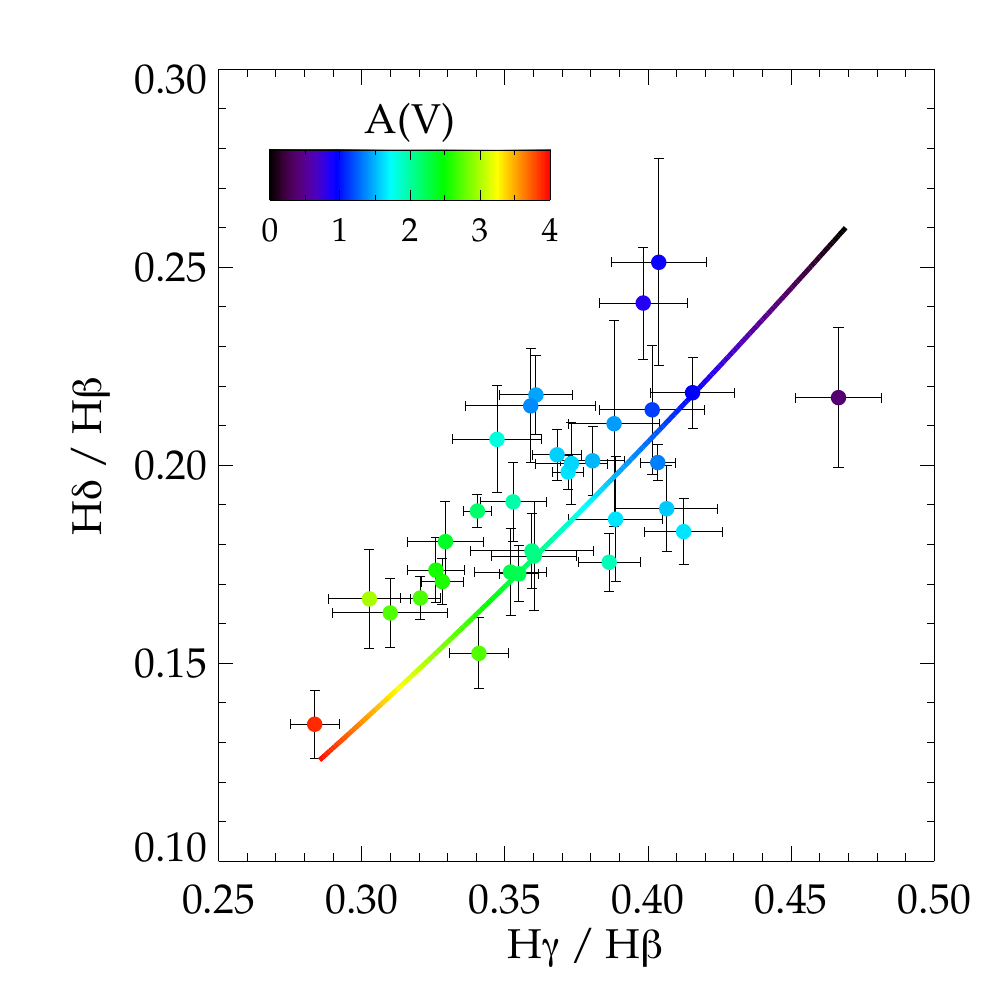}
\caption{
\label{fig:dust}
Balmer emission line ratios and best-fit extinction A(V). Multiple Balmer lines generally give consistent results. The solid colored line shows expected Balmer decrements as a function of A(V) for our adopted extinction curve with R$_{\mathrm V} = 4.05$. Different values of R$_{\mathrm V}$ result in an indistinguishable curve (given the measurement uncertainties), with A(V) varying in proportion to R$_{\mathrm V}$ in such a way that the color excess E(B$-$V)~$=$~A(V)/R$_{\mathrm V}$ is robust.
}
\end{figure}

Accurate correction for reddening is essential for the purposes of this study. We parameterize the reddening as a \cite{Cardelli1989} extinction curve with R$_{\mathrm V} = 4.05$, and determine the extinction and its uncertainty from a simultaneous fit to the Balmer lines \Hb, \Hg, and \Hd. Different Balmer line ratios generally give consistent results, agreeing within the formal $1\sigma$ uncertainty in 62\% of the sample. We do not use lines of higher order than \Hd\ because they suffer from non-negligible contamination (by \Neiii$\lambda$3968 and Ca H in the case of H$\epsilon$, and He~{\sc i}~$\lambda$3889 in the case of H$\zeta$) and increased sensitivity to stellar Balmer absorption. Balmer decrements and best-fit extinction A(V) for the $z\simeq0.8$ sample are shown in Figure~\ref{fig:dust}. All emission lines are corrected for the best fit extinction, and its uncertainty is propagated throughout the following analysis.
Reddening-correcting emission line fluxes are given in Table~\ref{tab:fluxes} along with best fit extinctions for each galaxy in the \Te\ sample.

\begin{deluxetable*}{lccccccccc}
\tablecolumns{5}
\tablewidth{0pt}
\tablecaption{De-reddened emission line fluxes\label{tab:fluxes}}
\tablehead{
\colhead{ID} & \colhead{$z$} & \colhead{\Oiii$\lambda$4959} & \colhead{\Hb} & \colhead{\Oiii$\lambda$4363} & \colhead{\Neiii$\lambda$3869} & \colhead{\Oii$\lambda\lambda$3727,3729} & \colhead{$\frac{\mathrm{[O~\textsc{ii}]}\lambda3729}{\mathrm{[O~\textsc{ii}]}\lambda3727}$} & \colhead{SNR(\Hb)\tablenotemark{1}} & \colhead{A(V)} \\
}
\medskip
\startdata
 1 & 0.771 & 2.11$\pm$0.05 & 1.00 & 0.092$\pm$0.015 & 0.528$\pm$0.040 & 1.48$\pm$0.11 & 1.30$\pm$0.06 &  44 & 3.01$\pm$0.28  \\
 2 & 0.819 & 1.42$\pm$0.04 & 1.00 & 0.081$\pm$0.014 & 0.340$\pm$0.027 & 1.82$\pm$0.13 & 1.31$\pm$0.04 &  44 & 1.66$\pm$0.26  \\
 3 & 0.746 & 2.07$\pm$0.02 & 1.00 & 0.115$\pm$0.006 & 0.540$\pm$0.012 & 0.96$\pm$0.02 & 1.25$\pm$0.03 & 114 & 1.65$\pm$0.08  \\
 4 & 0.747 & 1.05$\pm$0.02 & 1.00 & 0.018$\pm$0.009 & 0.229$\pm$0.010 & 2.65$\pm$0.09 & 1.10$\pm$0.02 & 104 & 2.58$\pm$0.12  \\
 5 & 0.762 & 1.56$\pm$0.03 & 1.00 & 0.043$\pm$0.010 & 0.358$\pm$0.019 & 2.70$\pm$0.13 & 1.34$\pm$0.02 &  73 & 1.46$\pm$0.18  \\
 6 & 0.780 & 0.93$\pm$0.02 & 1.00 & 0.025$\pm$0.005 & 0.219$\pm$0.010 & 2.37$\pm$0.08 & 1.44$\pm$0.03 &  83 & 2.72$\pm$0.12  \\
 7 & 0.771 & 2.08$\pm$0.03 & 1.00 & 0.067$\pm$0.008 & 0.483$\pm$0.017 & 1.66$\pm$0.06 & 1.25$\pm$0.02 &  93 & 1.60$\pm$0.12  \\
 8 & 0.766 & 1.50$\pm$0.05 & 1.00 & 0.028$\pm$0.013 & 0.320$\pm$0.032 & 2.54$\pm$0.20 & 1.54$\pm$0.04 &  47 & 1.44$\pm$0.29  \\
 9 & 0.762 & 1.28$\pm$0.02 & 1.00 & 0.032$\pm$0.011 & 0.335$\pm$0.021 & 2.67$\pm$0.15 & 1.37$\pm$0.03 &  78 & 2.37$\pm$0.22  \\
10 & 0.841 & 1.65$\pm$0.04 & 1.00 & 0.063$\pm$0.014 & 0.415$\pm$0.027 & 1.90$\pm$0.12 & 1.48$\pm$0.04 &  56 & 1.57$\pm$0.23  \\
11 & 0.767 & 2.01$\pm$0.04 & 1.00 & 0.026$\pm$0.019 & 0.469$\pm$0.025 & 1.85$\pm$0.09 & 1.33$\pm$0.04 &  71 & 1.99$\pm$0.18  \\
12 & 0.794 & 1.74$\pm$0.03 & 1.00 & 0.051$\pm$0.009 & 0.379$\pm$0.018 & 1.79$\pm$0.08 & 1.34$\pm$0.03 &  78 & 1.52$\pm$0.16  \\
13 & 0.798 & 1.91$\pm$0.05 & 1.00 & 0.095$\pm$0.014 & 0.434$\pm$0.031 & 1.66$\pm$0.11 & 1.40$\pm$0.05 &  44 & 1.82$\pm$0.25  \\
14 & 0.842 & 1.75$\pm$0.04 & 1.00 & 0.090$\pm$0.011 & 0.421$\pm$0.026 & 1.38$\pm$0.08 & 1.45$\pm$0.06 &  53 & 0.85$\pm$0.21  \\
15 & 0.796 & 0.84$\pm$0.01 & 1.00 & 0.007$\pm$0.005 & 0.160$\pm$0.009 & 2.15$\pm$0.07 & 1.03$\pm$0.01 & 139 & 2.24$\pm$0.12  \\
16 & 0.748 & 1.80$\pm$0.04 & 1.00 & 0.151$\pm$0.019 & 0.455$\pm$0.025 & 1.53$\pm$0.08 & 1.38$\pm$0.05 &  62 & 1.61$\pm$0.18  \\
17 & 0.738 & 1.53$\pm$0.04 & 1.00 & 0.092$\pm$0.012 & 0.520$\pm$0.027 & 1.66$\pm$0.08 & 1.39$\pm$0.04 &  54 & 2.55$\pm$0.17  \\
18 & 0.793 & 1.65$\pm$0.04 & 1.00 & 0.064$\pm$0.016 & 0.365$\pm$0.032 & 2.06$\pm$0.15 & 1.35$\pm$0.03 &  52 & 1.16$\pm$0.26  \\
19 & 0.793 & 1.85$\pm$0.01 & 1.00 & 0.073$\pm$0.005 & 0.435$\pm$0.010 & 1.93$\pm$0.04 & 1.42$\pm$0.02 & 169 & 2.18$\pm$0.08  \\
20 & 0.755 & 1.82$\pm$0.03 & 1.00 & 0.119$\pm$0.010 & 0.434$\pm$0.019 & 0.93$\pm$0.04 & 1.24$\pm$0.05 &  87 & 1.94$\pm$0.15  \\
21 & 0.788 & 1.63$\pm$0.04 & 1.00 & 0.052$\pm$0.015 & 0.426$\pm$0.034 & 1.74$\pm$0.12 & 1.23$\pm$0.05 &  54 & 2.11$\pm$0.25  \\
22 & 0.856 & 2.17$\pm$0.05 & 1.00 & 0.064$\pm$0.012 & 0.330$\pm$0.020 & 1.20$\pm$0.07 & 1.25$\pm$0.04 &  58 & 0.40$\pm$0.21  \\
23 & 0.732 & 1.44$\pm$0.02 & 1.00 & 0.044$\pm$0.010 & 0.407$\pm$0.024 & 2.56$\pm$0.13 & 1.22$\pm$0.03 &  85 & 3.88$\pm$0.19  \\
24 & 0.802 & 1.44$\pm$0.05 & 1.00 & 0.011$\pm$0.009 & 0.338$\pm$0.028 & 2.08$\pm$0.17 & 1.35$\pm$0.03 &  33 & 1.39$\pm$0.29  \\
25 & 0.766 & 2.01$\pm$0.06 & 1.00 & 0.058$\pm$0.013 & 0.408$\pm$0.034 & 1.69$\pm$0.13 & 1.35$\pm$0.04 &  51 & 0.94$\pm$0.28  \\
26 & 0.721 & 1.78$\pm$0.03 & 1.00 & 0.075$\pm$0.010 & 0.407$\pm$0.019 & 1.28$\pm$0.06 & 1.19$\pm$0.03 &  78 & 0.96$\pm$0.17  \\
27 & 0.774 & 1.64$\pm$0.04 & 1.00 & 0.070$\pm$0.011 & 0.353$\pm$0.026 & 2.23$\pm$0.14 & 1.30$\pm$0.03 &  42 & 2.27$\pm$0.21  \\
28 & 0.731 & 1.99$\pm$0.02 & 1.00 & 0.063$\pm$0.004 & 0.492$\pm$0.012 & 1.41$\pm$0.03 & 1.25$\pm$0.02 & 133 & 1.34$\pm$0.08  \\
29 & 0.751 & 1.70$\pm$0.03 & 1.00 & 0.060$\pm$0.013 & 0.363$\pm$0.025 & 1.93$\pm$0.12 & 1.38$\pm$0.04 &  78 & 2.08$\pm$0.24  \\
30 & 0.833 & 1.39$\pm$0.02 & 1.00 & 0.023$\pm$0.009 & 0.361$\pm$0.018 & 2.39$\pm$0.11 & 1.26$\pm$0.02 &  84 & 1.65$\pm$0.17  \\
31 & 0.749 & 1.98$\pm$0.04 & 1.00 & 0.056$\pm$0.017 & 0.442$\pm$0.025 & 1.95$\pm$0.10 & 1.44$\pm$0.04 &  63 & 2.74$\pm$0.19  \\
32 & 0.751 & 1.85$\pm$0.04 & 1.00 & 0.063$\pm$0.013 & 0.418$\pm$0.029 & 1.73$\pm$0.12 & 1.32$\pm$0.05 &  69 & 2.73$\pm$0.25  \\
\enddata
\tablenotetext{1}{Signal to noise ratio of \Hb\ emission.}
\end{deluxetable*}

\begin{figure}
\includegraphics[width=\columnwidth]{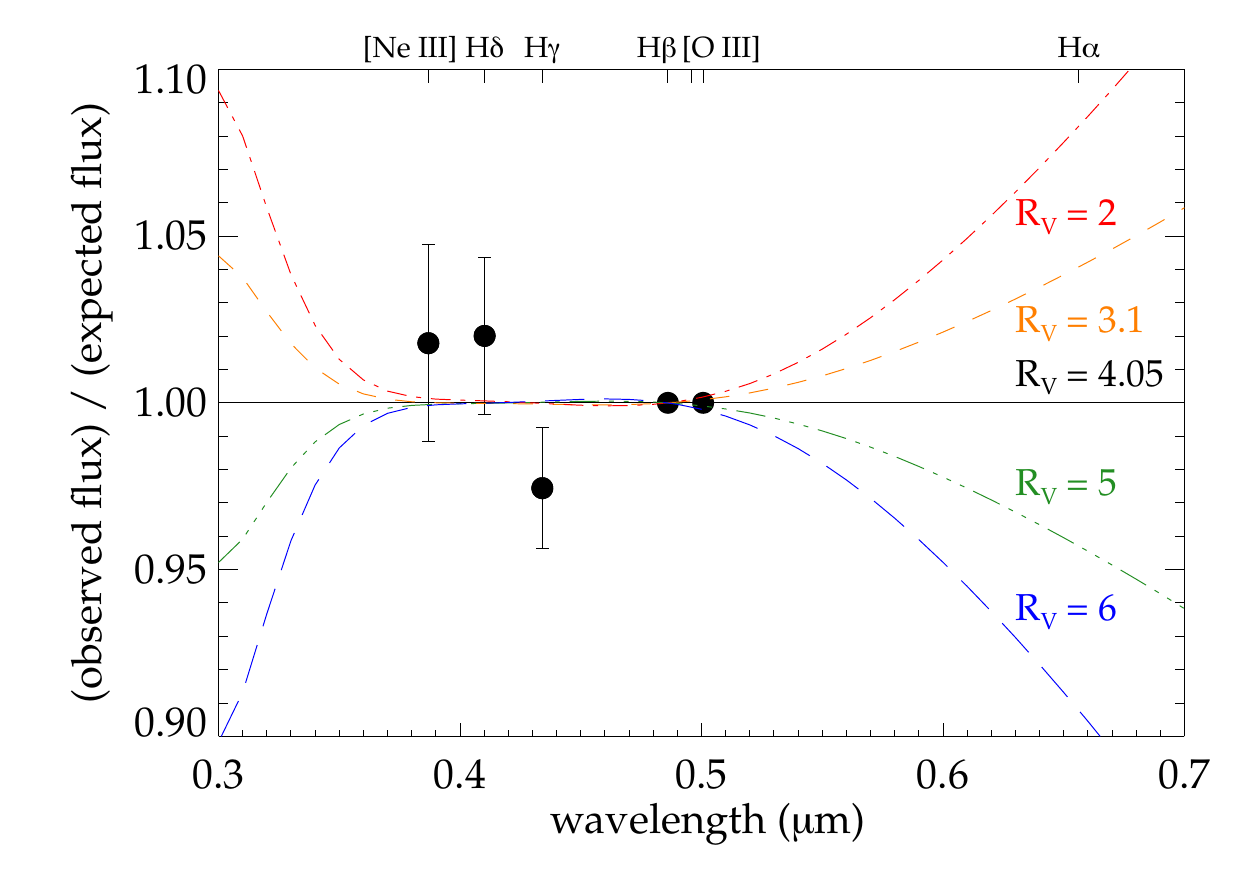}
\caption{
\label{fig:rv}
Emission line flux ratios compared to various \cite{Cardelli1989} extinction curves, demonstrating a degeneracy in R$_{\mathrm V}$. All values are relative to a fiducial curve obtained by fitting the sample mean Balmer emission line ratios with R$_{\mathrm V} = 4.05$. Data points show sample mean Balmer fluxes relative to \Hb\ (assuming intrinsic Balmer decrements from case B recombination: \Hg/\Hb~$=0.47$, \Hd/\Hb~$=0.26$) and mean \Neiii\ relative to \Oiii\ (assuming intrinsic \Neiii/\Oiii~$=0.077$, the mean de-reddened value from the local comparison sample). We additionally fit the Balmer line ratios using extinction curves with various R$_{\mathrm V}$ (dashed lines). We note that R$_{\mathrm V} = 3.1$ and 4.05 are typically assumed for the Milky Way \citep{Cardelli1989} and for starburst galaxies \citep[e.g.,][]{Calzetti2000}, respectively. 
The data are in excellent agreement and cannot constrain R$_{\mathrm V}$ due to the short wavelength range probed, although different R$_{\mathrm V}$ curves diverge rapidly at longer and shorter wavelengths. 
Future measurements of \Ha\ or other features beyond the present wavelength range could therefore be used to determine R$_{\mathrm V}$.
}
\end{figure}

Balmer line fluxes provide tight constraints on the color excess E(B$-$V)~$=$~A(V)/R$_{\mathrm V}$ for all galaxies in the \Te\ sample. However, we cannot meaningfully constrain R$_{\mathrm V}$ due to the limited rest-frame wavelength range of the DEEP2 spectra. Consequently the total extinction A(V) is uncertain, but this does not affect our results since we are concerned only with line ratios. Fortunately all de-reddened emission line ratios used in this work are insensitive to R$_{\mathrm V}$ (Figure~\ref{fig:rv}) and are therefore robust. Future observations of \Ha\ or Paschen series lines could be used to measure R$_{\mathrm V}$ and provide accurate de-reddened emission line luminosities.

Since many of the line ratios of interest are sensitive to reddening corrections (such as \Oiii/\Oii), we have carefully considered possible associated systematic errors. One potential source of error is the correction for stellar Balmer absorption, which corresponds to an average increase of 1\% in \Hg/\Hb\ and 3\% in \Hd/\Hb\ for our sample. This is small compared to the total uncertainty in the line ratios (Figure~\ref{fig:dust}) and does not introduce significant systematic errors. Even for an extreme case where the true stellar absorption has been underestimated by a factor of 2, the ratio of \Oiii$\lambda$5007/\Oii$\lambda\lambda$3727,3729 would be within 0.02 dex of the true value (with smaller corrections for other line ratios). We also examined the effect of differential atmospheric refraction which may cause increasing slit losses at bluer wavelengths. We use the prescriptions of \cite{Filippenko1982} to calculate slit loss as a function of wavelength for the median seeing, airmass, and slit position angle of the DEIMOS observations. The expected effect is that the total extinction is overestimated by $\simeq$0.01 dex, with \Oii\ and \Neiii\ emission line fluxes underestimated by only $\lesssim0.01$ dex compared to the other emission lines used in this work. In summary, systematic errors in the de-reddened line ratios are expected to be $\lesssim$0.02 dex. Despite this small uncertainty we strive to use reddening-independent line ratios (i.e., with close wavelength spacing such as in Figure~\ref{fig:o3bias}) wherever possible.

\subsection{Electron temperature and density}

\begin{figure}
\centerline{\includegraphics[width=0.9\columnwidth]{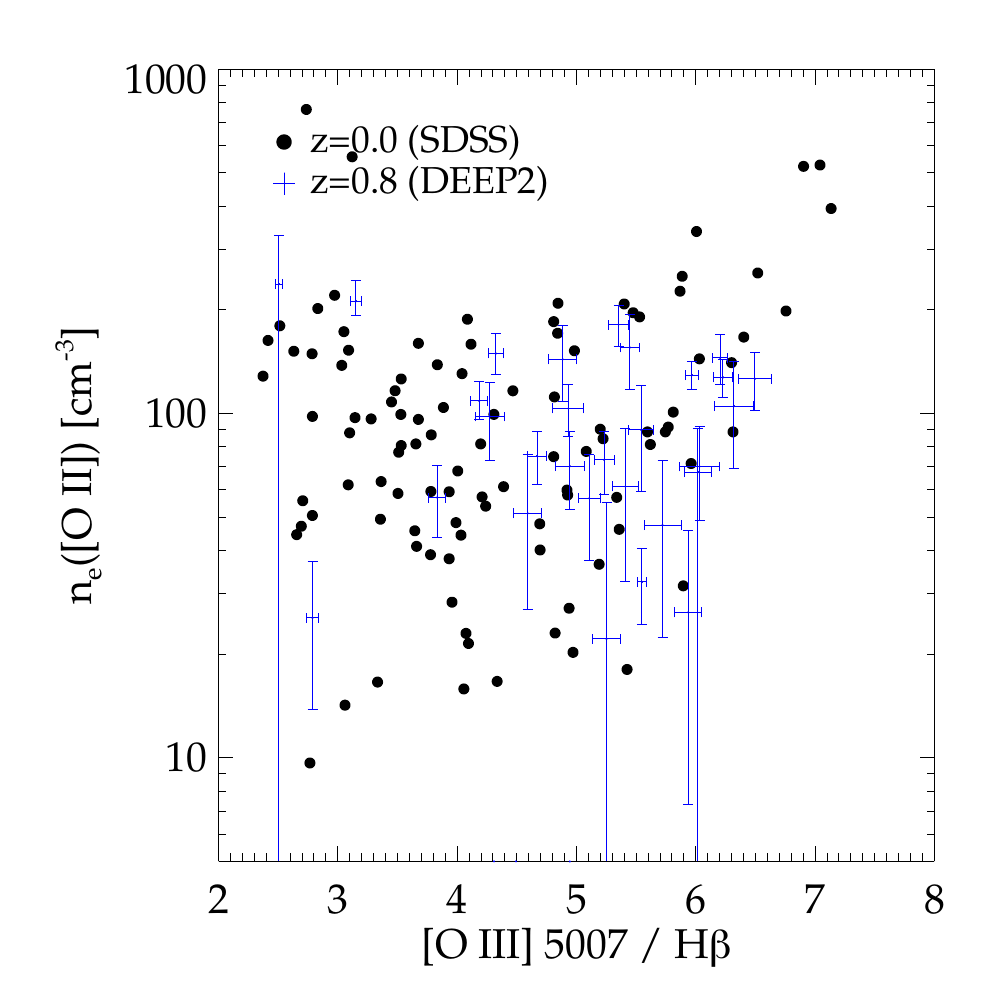}}
\centerline{\includegraphics[width=0.9\columnwidth]{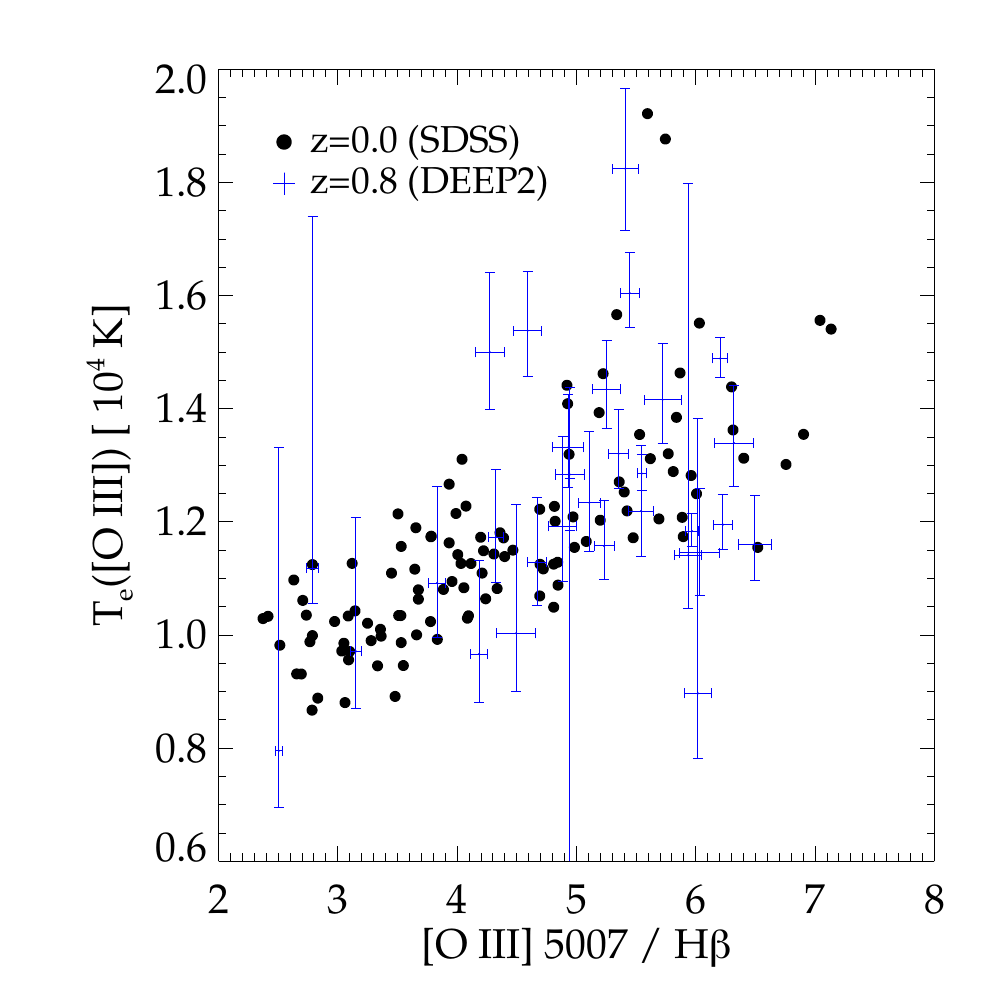}}
\caption{
\label{fig:nt}
Nebular electron density \Ne\ and temperature \Te\ as a function of \Oiii$/$\Hb. There is no evolution in these properties: galaxies at $z\simeq0.8$ have densities and temperatures which are typical of local galaxies with similar strong line ratios. 
Densities are derived from the \Sii$\lambda\lambda$6717,6731 doublet for the $z=0$ comparison sample and from the \Oii$\lambda\lambda$3727,3729 doublet for the $z\simeq0.8$ sample. Temperatures are derived from the \ROiii\ ratio for both cases.
}
\end{figure}

Nebular electron temperatures and densities are derived using the {\sc IRAF nebular.temdens} package. We first estimate temperatures \Te\ from the \ROiii\ ratio assuming an electron density \Ne$= 100$ cm$^{-3}$, and then calculate \Ne\ from the \Oii$\lambda\lambda$3727,3729 doublet ratio using the derived \Te. This process is iterated and results in no significant change compared to uncertainty in the flux ratios. Importantly, we find values of \Ne $< 400$ cm$^{-3}$ and \Te $<2\times10^4$ K in all cases, as shown in Figure~\ref{fig:nt}. In this regime \Te\ is insensitive to \Ne\ ($<$50 K variation). Density is weakly sensitive to \Te\ in this regime and we account for this in the uncertainty of \Ne, although the effect on derived abundances is minimal ($\lesssim$0.01 dex). The \Te\ values we derive differ by $<$300 K from the analytic method used by \cite{Izotov2006}, resulting in $<$0.01 dex difference in the derived oxygen abundance. Figure~\ref{fig:nt} shows \Ne\ and \Te\ of our sample and the local comparison sample. Notably, we find no evidence of evolution in these physical properties for galaxies with similar ratios of strong emission lines.

We are unable to measure the singly ionized oxygen temperature from DEEP2 spectra. Instead we estimate \Te(\Oii) from measurements of \Te(\Oiii) using the same method as \cite{Izotov2006} in order to facilitate direct comparison of the results. We measure a scatter about the best-fit linear \Te(\Oii)--\Te(\Oiii) relation of $\sigma$(\Te(\Oii)) $=1030$~K from the data presented by \cite{Pilyugin2010}. While this overestimates the true variation in \Te(\Oii) at fixed \Te(\Oiii), we include a conservative uncertainty of $10^3$ K in our estimates of \Te(\Oii). This is added in quadrature with uncertainty arising from measurement error. We caution that various prescriptions for the \Te(\Oii)--\Te(\Oiii) relation exist in the literature, and derived metallicities are moderately sensitive to the adopted relation. For example, the measurements of \cite{Pilyugin2010} suggest values of \Te(\Oii) which are $\simeq$10$^3$ K lower on average than those we adopt; this would result in a $\simeq$0.02 dex increase in oxygen abundances for our sample.

\subsection{Metallicity}

\begin{figure}
\centerline{\includegraphics[width=\columnwidth]{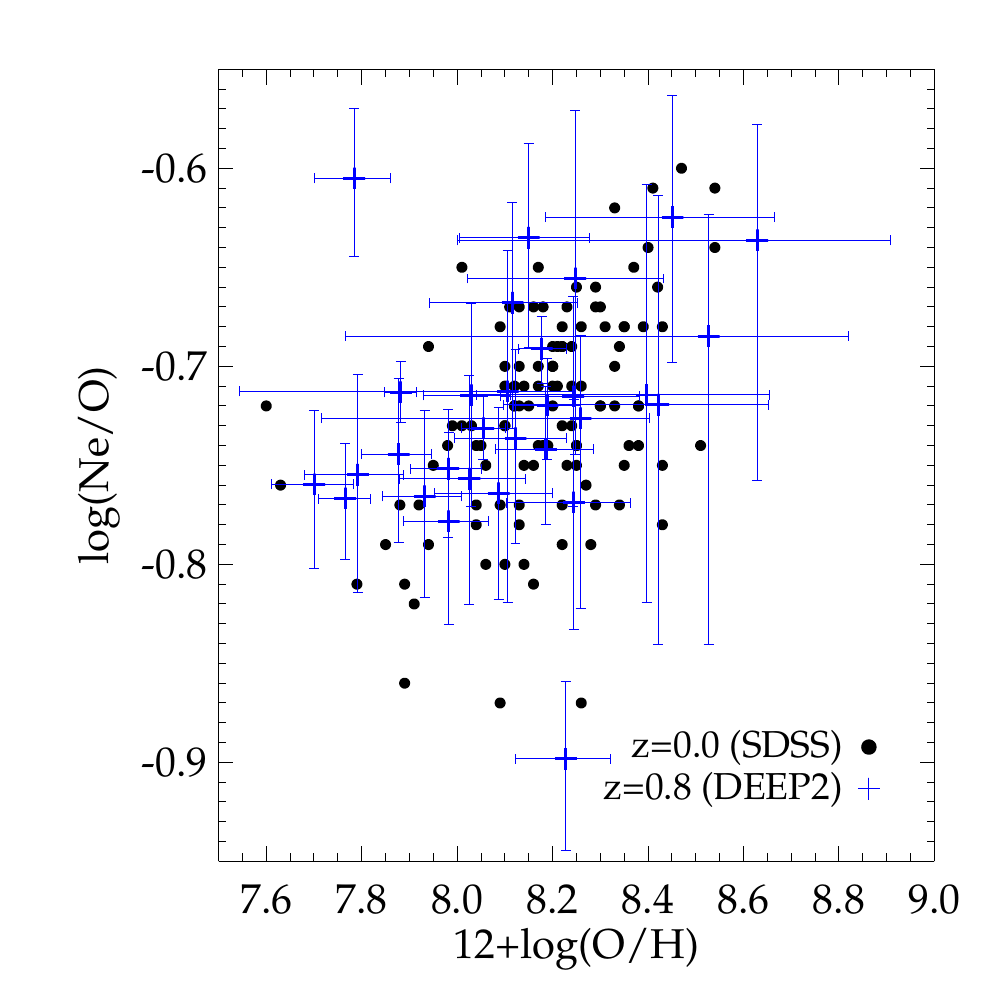}}
\caption{
\label{fig:neon}
Neon and oxygen abundances. The Ne/O abundance ratio at $z\simeq0.8$ is consistent with local galaxies; the median value is $\log{\mathrm{Ne/O}} = -0.72\pm0.01$ for both samples.
For reference, estimates of the solar value are typically $\log{\mathrm{Ne/O}_{\odot}} = -0.8$ to $-0.65$ \citep{Asplund2009}.
}
\end{figure}

Ionic abundances are derived from the density, temperature, and de-reddened line ratios of each galaxy in our sample. For consistency we adopt the analytical formulae used by \cite{Izotov2006} to derive abundances of O$^+$/H$^+$, O$^{2+}$/H$^+$, and Ne$^{2+}$/H$^+$. 
We calculate the total oxygen abundance as
$$\mathrm{\frac{O}{H} = \frac{O^+}{H^+} + \frac{O^{2+}}{H^+}}.$$
No correction is made for higher ions, which are expected to account for $<$1\% in all but three galaxies in our sample based on their O$^+$/O$^{2+}$ ratios \citep{Izotov2006}. 
We have also calculated abundances using the IRAF {\sc nebular.ionic} package, and we find that IRAF produces lower metallicities by $\Delta\log{\mathrm{O/H}} \simeq 0.03$ dex. For neon, we convert from Ne$^{2+}$/H$^+$ to total Ne/H using the same method as \cite{Izotov2006}. O/H and Ne/O abundances for our sample are shown in Figure~\ref{fig:neon}. Relative abundances of neon and oxygen are measured with higher precision than O/H and Ne/H, since Ne/O is less sensitive to the electron temperature.

\subsection{Uncertainty}\label{sec:uncertainty}

We have been careful to assess and mitigate potential sources of systematic uncertainty in deriving physical quantities. Ne/O abundance ratio measurements (Figure~\ref{fig:neon}) provide an empirical estimate of the degree of both random and systematic uncertainty. Neon and oxygen are predominantly generated by the same nucleosynthetic processes, and consequently Ne/O abundance shows little variation in \Hii\ regions and planetary nebulae spanning a wide range of metallicity and other physical properties \citep[e.g.,][]{Izotov2011,Leisy2006}. The modest increase in Ne/O at higher O/H, amounting to a variation of $\Delta\log{\mathrm{Ne/O}}\simeq0.1$ dex across the samples considered here, is discussed further in Section~\ref{sec:neon}.

Both the DEEP2 and local comparison samples have a median abundance ratio of $\log{\mathrm{Ne/O}} = -0.72\pm0.01$. Such excellent agreement indicates that any systematic errors affecting the comparison between these two samples are limited to $\lesssim0.01$ dex. In particular, Ne/O is highly sensitive to estimates of reddening and differential refraction due to the relatively large wavelength separation of \Neiii$\lambda$3869 and \Oiii$\lambda\lambda$4959,5007. We infer that there is minimal uncertainty arising from these and other effects, or else that both samples are affected by similar amounts. Most importantly this confirms that direct comparison of the two samples yields reliable results. We caution that studies employing different methods are not necessarily comparable (e.g., abundances derived from IRAF's {\sc nebular.ionic} task differ systematically by $\simeq0.03$ dex as discussed above).

Additionally, 19 of the 32 galaxies in the DEEP2 sample are compatible with $\log{\mathrm{Ne/O}} = -0.72$ at the $1\sigma$ level, consistent with expectations for an approximately constant Ne/O abundance ratio. Allowing for a trend of increasing Ne/O with O/H, we find that 24/32 galaxies are consistent ($1\sigma$) with the best-fit linear relation, also in agreement with statistical expectations. This sanity check confirms that uncertainties adopted for our sample are similar to the true measurement error.

\section{Strong-line abundance diagnostics}\label{sec:diagnostics}

We construct a set of diagnostic relations between oxygen abundance derived in Section~\ref{sec:properties}, expressed as $12+\log{\mathrm{O/H}}$, and the following strong emission line ratios: \Neiii/\Oii, \Oiii/\Oii, \Oiii/\Hb, \Oii/\Hb, and $\mathrm{R}_{23} = \frac{\mathrm{[O~\textsc{ii}] + [O~\textsc{iii}]}\lambda\lambda4959,5007}{\mathrm{H}\beta}$. Since we find no evolution in these relations (Section~\ref{sec:evolution}), we derive them from the local comparison sample. Independent calibrations of the $z\simeq0.8$ sample are fully consistent but with larger uncertainty. A calibration of the combined data sets also gives consistent results and does not substantially improve the precision.

We anchor the metallicity calibrations to fits of \Neiii/\Oii\ and \Oiii/\Oii\ as a function of metallicity. These two relations are examined first because they are known to be monotonic, at least for local galaxies \citep[e.g.,][]{Maiolino2008}. We fit the data using a functional form
\begin{equation}\label{eq:linear}
\log{\mathrm{R}} = c_0 + c_1 x,
\end{equation}
where $\mathrm{R}$ is the line ratio and $x = 12 + \log{\mathrm{O/H}}$. The resulting fits have reduced $\chi^2_{\nu}$ values larger than one, which we attribute to intrinsic scatter in these relations. Assuming that the estimated uncertainties are accurate, we calculate the intrinsic scatter $\sigma_{\mathrm{int}}$ by solving for the expectation that
\begin{equation}\label{eq:scatter}
\chi^2_{\nu} = \frac{1}{N} \sum \frac{\Delta\log{\mathrm{R}}}{\sigma^2_{\mathrm{int}} + \sigma^2_{\mathrm{meas}}} = 1,
\end{equation}
where $\sigma_{\mathrm{meas}}$ is the measurement uncertainty and $N$ is the number of degrees of freedom in the fit.

The other line ratios exhibit significant higher-order trends and we therefore adopt a second-order polynomial relation
\begin{equation}\label{eq:quadratic}
\log{\mathrm{R}} = c_0 + c_1 x + c_2 x^2.
\end{equation}
However, fitting $\log{\mathrm{R}}$ to $\log{\mathrm{O/H}}$ directly generally fails to capture the non-linear behavior, likely because of limited dynamic range and relatively large uncertainty in metallicity. Including higher order polynomial terms only exacerbates this problem. Therefore, we obtain metallicity calibrations by fitting the line ratios to
\begin{equation}
\log{\mathrm{R}} = a_0 + a_1 y + a_2 y^2
\end{equation}
with $y = \log{\mathrm{[Ne~\textsc{iii}]/[O~\textsc{ii}]}}$. (Using $y = \log{\mathrm{[O~\textsc{iii}]/[O~\textsc{ii}]}}$ gives consistent results, but is subject to larger uncertainty arising from reddening corrections.) This provides improved fits since these strong emission line ratios are measured with much better precision than oxygen abundance. Combining these fits with the calibrations derived from Equation~\ref{eq:linear} gives a relation in the form of Equation~\ref{eq:quadratic}. Intrinsic scatter is derived from Equation~\ref{eq:scatter} in the same way as the other diagnostics.

\begin{figure*}
\centerline{
\includegraphics[width=0.7\columnwidth]{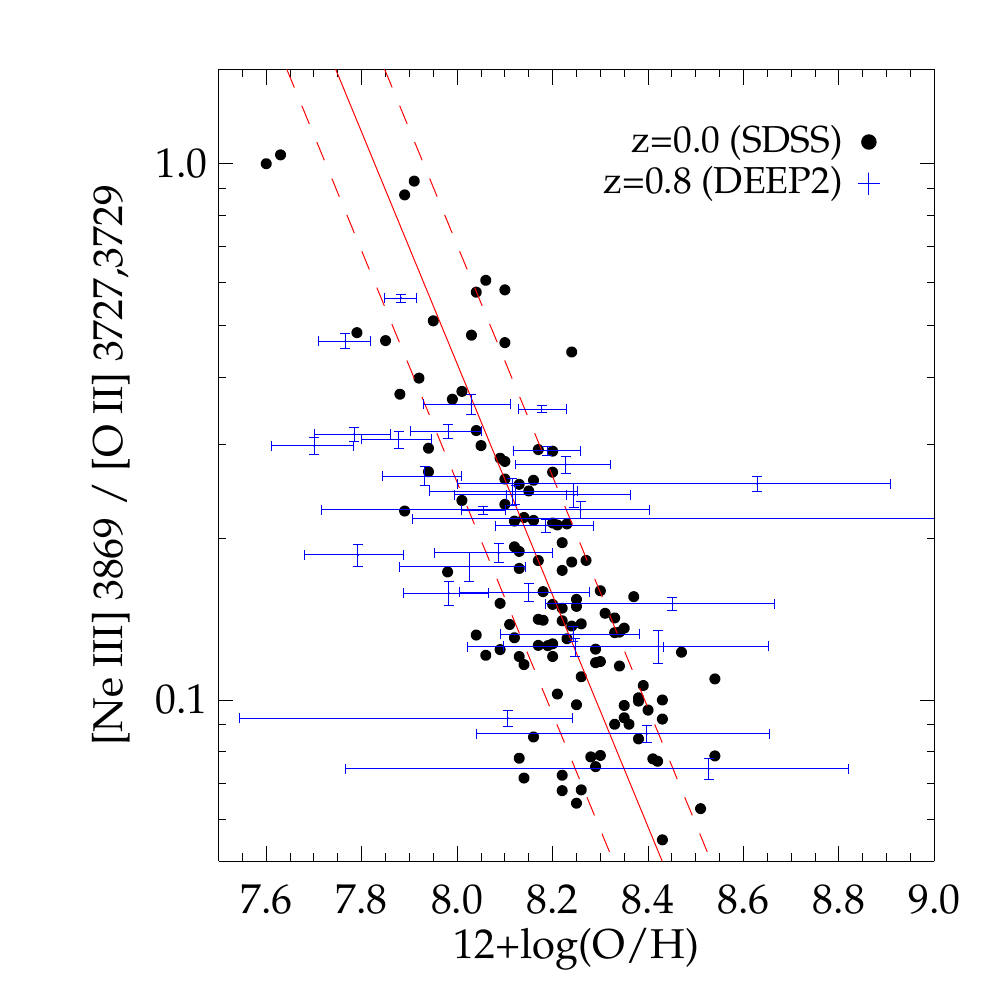}
\includegraphics[width=0.7\columnwidth]{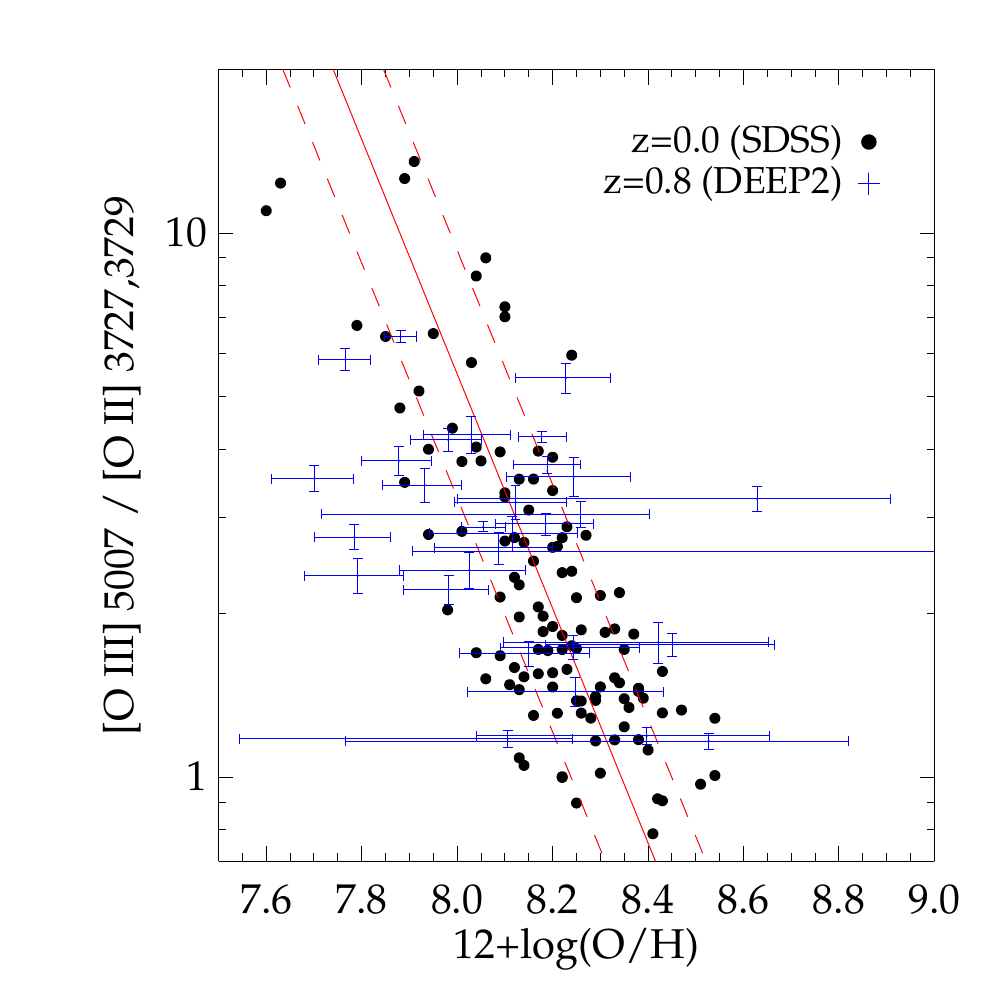}
\includegraphics[width=0.7\columnwidth]{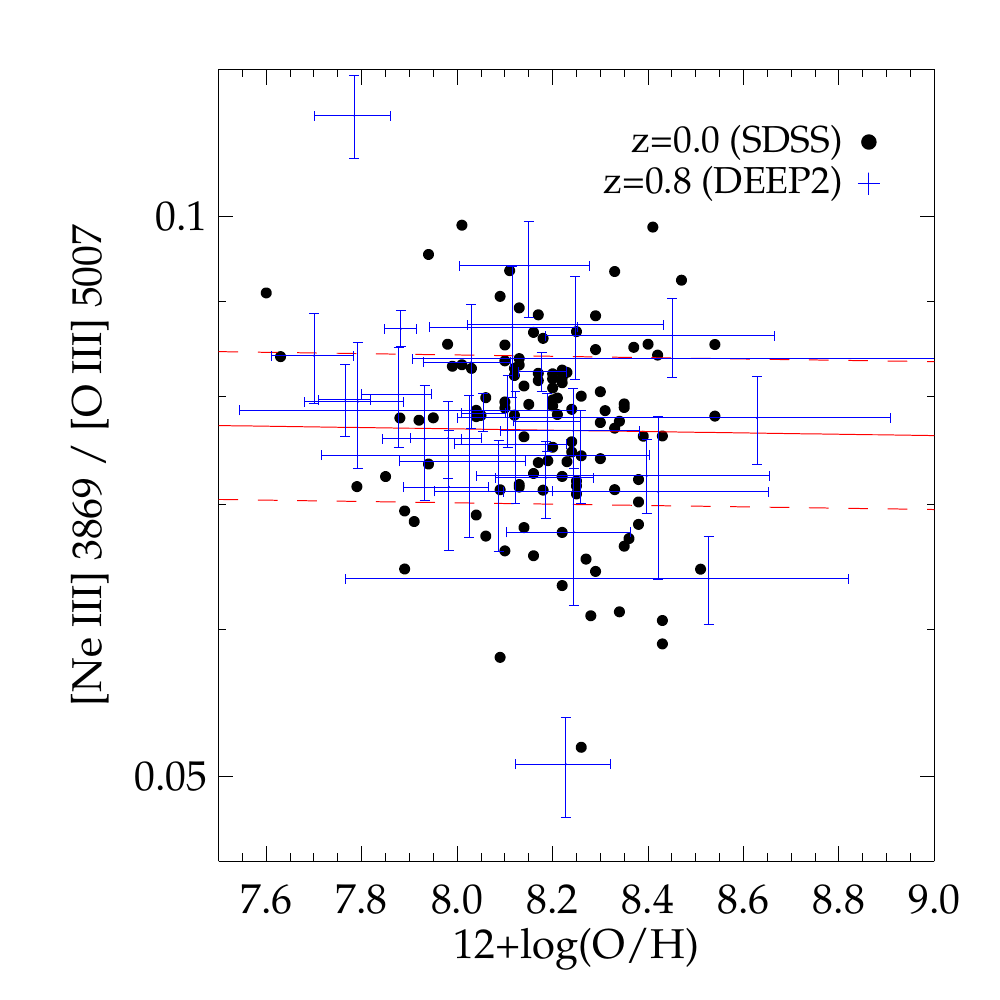}
}
\centerline{
\includegraphics[width=0.7\columnwidth]{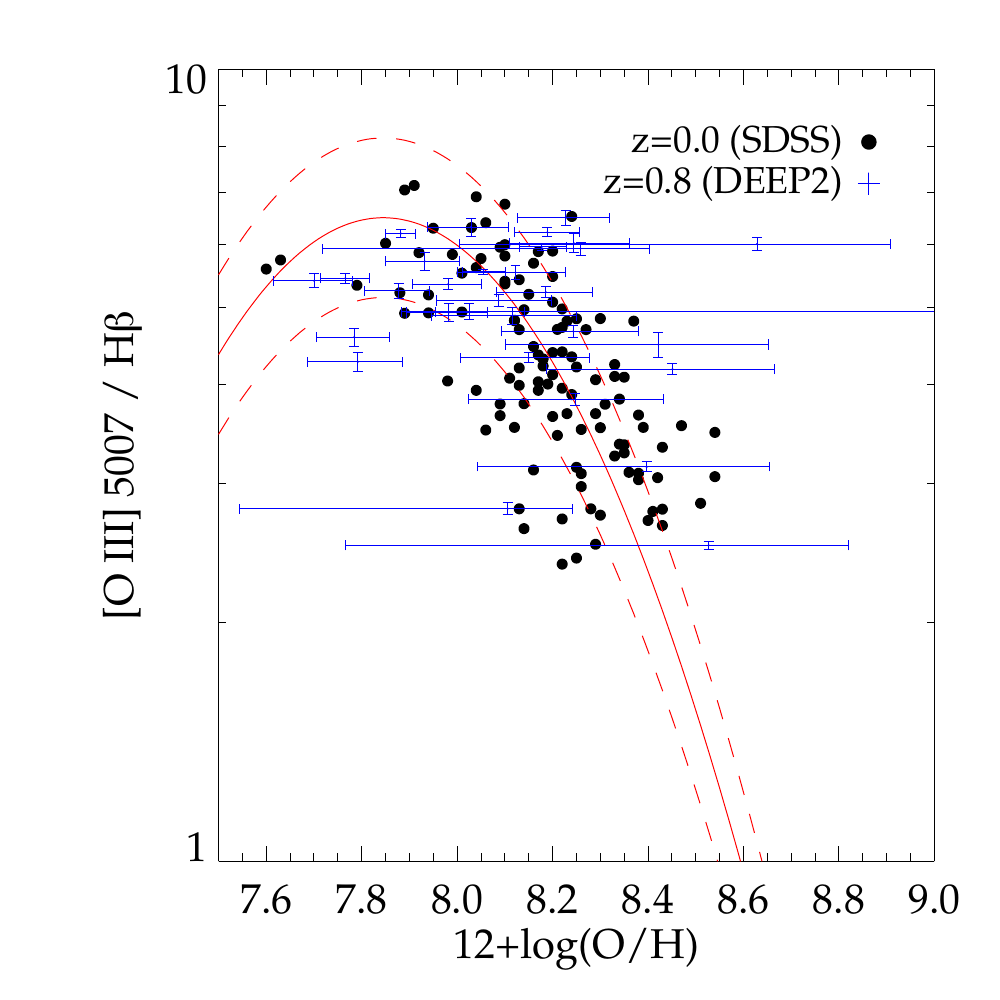}
\includegraphics[width=0.7\columnwidth]{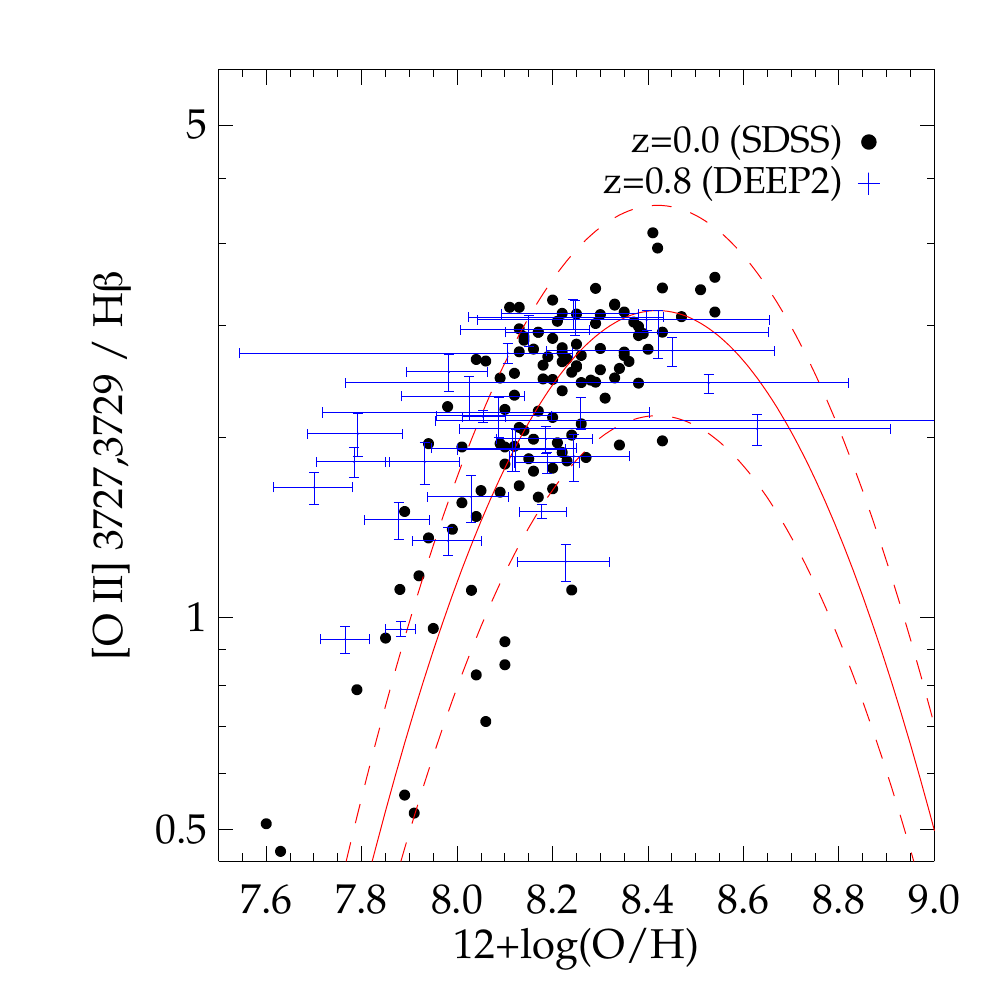}
\includegraphics[width=0.7\columnwidth]{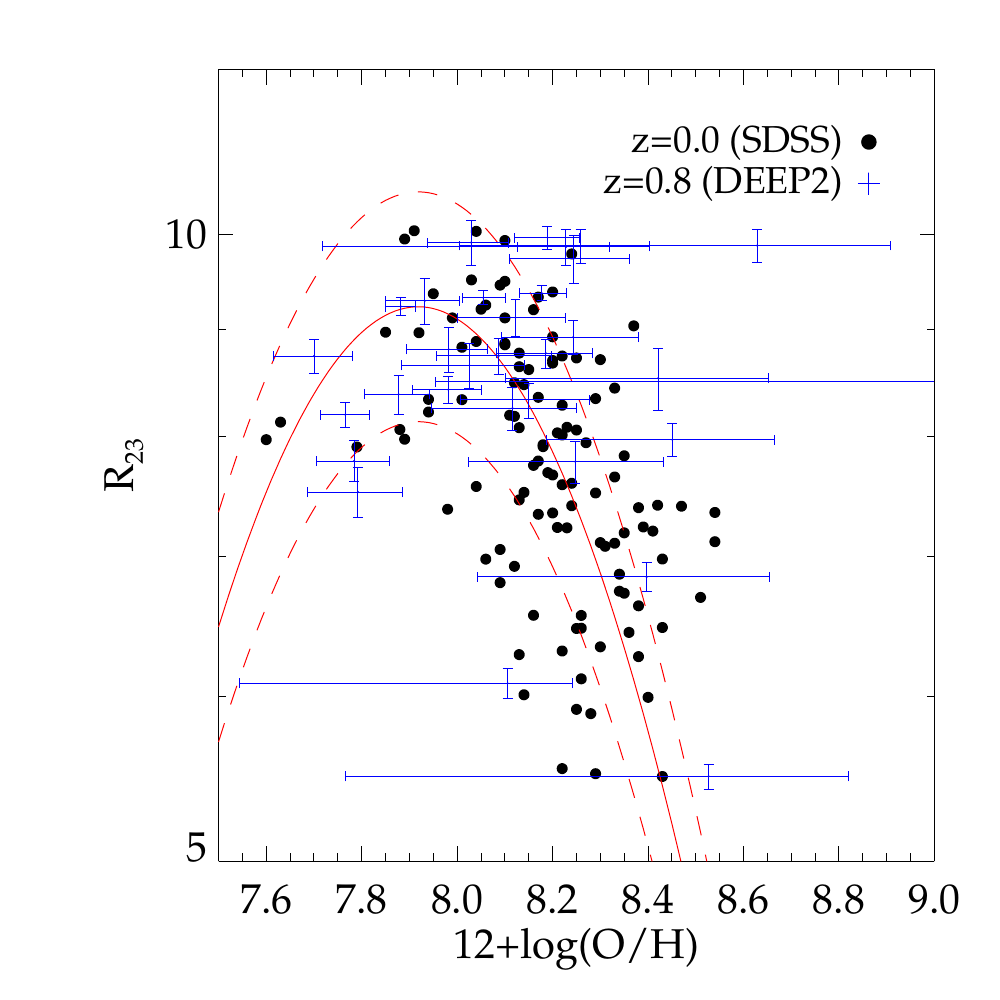}
}
\caption{
\label{fig:offset}
Strong emission line ratios as a function of metallicity. Solid lines show best fit relations derived for the $z\simeq0$ sample, and dashed lines show the intrinsic scatter in each relation.
Data at $z\simeq0.8$ are consistent with the $z\simeq0$ comparison sample to within the 1$-\sigma$ measurement precision of $\Delta \log{\mathrm{O/H}} \lesssim 0.03$ dex.
}
\end{figure*}

Best-fit relations for each strong line abundance diagnostic are shown in Figure~\ref{fig:offset}. Table~\ref{tab:diagnostics} lists the corresponding fit coefficients and intrinsic scatter. 
We also include the relation between metallicity and \Neiii/\Oiii\ obtained by combining the \Neiii/\Oii\ and \Oiii/\Oii\ calibrations (rather than a direct fit), as well as the O2Ne3 calibration introduced by \cite{Perez-Montero2007}:
$$\mathrm{O2Ne3} = \frac{\mathrm{[O~\textsc{ii}]} + 17.06 \mathrm{[Ne~\textsc{iii}]}}{\mathrm{H}\delta} \approx \mathrm{R}_{23} \frac{\mathrm{H}\beta}{\mathrm{H}\delta}, $$
which we obtain from multiplying the $\mathrm{R_{23}}$ calibration by the Balmer line ratio (\Hd/\Hb~$=0.260$). The value 17.06 represents the mean \Oiii$\lambda\lambda$4959,5007$/$\Neiii\ ratio. In both of these cases we calculate the intrinsic scatter from Equation~\ref{eq:scatter}.
While the values in Table~\ref{tab:diagnostics} are derived from the local comparison sample, results from the $z\simeq0.8$ \Te\ sample are consistent with the same coefficients and scatter. Higher order polynomial terms do not significantly improve the fit quality or derived scatter compared to the adopted results from Equations~\ref{eq:linear} and \ref{eq:quadratic}. We caution that these diagnostics are valid only within the range of abundances and line ratios probed by the analyzed data.

\begin{deluxetable}{lrrrc}
\tablecolumns{5}
\tablewidth{0pt}
\tablecaption{Strong-line abundance diagnostic coefficients\label{tab:diagnostics}}
\tablehead{
\colhead{$\mathrm{R}$} & \colhead{$c_0$} & \colhead{$c_1$} & \colhead{$c_2$} & \colhead{$\sigma_{\mathrm{int}}$\tablenotemark{1} (dex)} \\
}
\medskip
\startdata
\Neiii/\Oii  &  16.8974  &  -2.1588  &    &  0.22  \\
\Oiii/\Oii  &  17.9828  &  -2.1552  &    &  0.23  \\
\Neiii/\Oiii &  -1.0854  &  -0.0036  &    &  0.04  \\
\Oiii/\Hb  &  -88.4378  &  22.7529  &  -1.4501  &  0.10  \\
\Oii/\Hb  &  -154.9571  &  36.9128  &  -2.1921  &  0.15  \\
$\mathrm{R_{23}}$  &  -54.1003  &  13.9083  &  -0.8782  &  0.06  \\
O2Ne3              &  -53.5153  &  13.9083  &  -0.8782  &  0.08  \\
\enddata
\tablenotetext{1}{Intrinsic scatter in $\log{\mathrm{R}}$ at fixed $\log{\mathrm{O/H}}$.}
\end{deluxetable}

\section{Results}\label{sec:results}

\subsection{No evolution in temperature and density}

Figure~\ref{fig:nt} demonstrates that \Te\ and \Ne\ do not evolve significantly between $z=0\rightarrow0.8$ at fixed emission line ratios. 
We quantify this with linear fits to the data and measure median offsets $\Delta\log{\mathrm{T_e}} = -0.013\pm0.014$ and $\Delta\log{\mathrm{n_e}} = -0.05\pm0.07$ in the DEEP2 sample at $z\simeq0.8$, relative to the local comparison sample at fixed \Oiii/\Hb. Similarly there are no significant differences in density or temperature relative to any other available line ratio, nor to the derived metallicity. \Oiii/\Hb\ is used here in order to address the possibility that systematically higher \Ne\ may cause elevated ratios of \Oiii/\Hb\ and \Nii/\Ha\ observed at high redshift \citep[e.g.,][]{Kewley2013,Shirazi2014}. We find no evidence for such an effect at $z=0.8$.

\subsection{No evolution in Ne/O abundance}\label{sec:neon}

The relative abundances of neon and oxygen are shown in Figure~\ref{fig:neon} and have been discussed briefly in Section~\ref{sec:uncertainty}. We now consider the positive correlation between Ne/O and O/H and its implications for high redshift galaxies. Large samples of local galaxies indicate a slope $\frac{\Delta\log{\mathrm{Ne/O}}}{\Delta\log{\mathrm{O/H}}} = 0.097\pm0.015$ \citep[Equation~30 of][]{Izotov2006}, consistent with the data at $z\simeq0.8$. We find no convincing evidence for evolution in either the slope or normalization of this relation: our \Te\ sample is offset by a median $\Delta\log{\mathrm{Ne/O}} = 0.011\pm0.008$ dex relative to the linear fit at $z=0$. The cause of increasing Ne/O with O/H is not entirely clear, however. 
\cite{Izotov2006,Izotov2011} attribute this relation to depletion of oxygen onto dust grains, implying $\sim$20\% of interstellar oxygen in the solid state for the most metal-rich galaxies in their sample. Alternatively, nucleosynthetic processes such as $\alpha$-capture onto oxygen or conversion of $^{14}$N$\rightarrow^{22}$Ne have been proposed to explain the variation in Ne/O (as discussed in, e.g., section 4.5 of \citealt{Leisy2006}).

Theoretical supernova yields can provide some insight into the effects of nucleosynthesis and dust depletion. For this purpose we integrated the yield tables of \cite{Nomoto2013} for core-collapse supernovae with a \cite{Salpeter1955} initial mass function truncated at 50 $\Msun$. (The IMF-integrated tables of \citealt{Kobayashi2006} give similar results). This gives an expected $\log{\mathrm{Ne/O}} = -0.71$ for progenitor stars with 1$/$5th solar metallicity, increasing to $\log{\mathrm{Ne/O}} = -0.67$ for 2$/$5th solar metallicity due to roughly equal effects of higher neon and lower oxygen yields. We note that the $^{22}$Ne production and hence $^{14}$N$\rightarrow^{22}$Ne reaction is negligible for these models. These predictions are in excellent agreement with the slope in Figure~\ref{fig:neon}, indicating that metallicity-dependent supernova yields can explain the trend. The small offset in predicted yields compared to the data at fixed metallicity is in the opposite direction expected from depletion of oxygen onto dust, hence the models suggest that depletion is not important. However we caution that this interpretation is subject to significant uncertainty in the theoretical yields.

We can independently test whether the dust depletion hypothesis is plausible by using nebular extinction (Section~\ref{sec:extinction}) as a proxy for dust content. Despite low measurement uncertainty there is no significant correlation between extinction and Ne/O in either the \Te or local comparison samples, in contrast to the $>$3$\sigma$ trend between O/H and Ne/O (Figure~\ref{fig:neon}). Therefore Ne/O appears to be more fundamentally related to O/H than to dust content. This supports nucleosynthetic processes rather than dust depletion as the cause of variation in Ne/O, although we caution that knowledge of the dust-to-gas mass fractions would be needed for robust conclusions. The trend of increasing Ne/Ar with O/H (as pointed out by, e.g., \citealt{Izotov2006}) further suggests that nucleosynthetic yields of Ne and/or Ar vary with overall metallicity since neither of these noble gases is affected by dust depletion.

\cite{Zeimann2015} have recently presented evidence for a $\sim$0.2 dex increase in \Neiii/\Oiii\ flux ratios at $z\simeq2$ compared to local galaxies, although its origin is unclear. In contrast, we find that line ratios at $z\simeq0.8$ are consistent with the locus of local star-forming galaxies (Figures~\ref{fig:o3bias}, \ref{fig:offset}). The \Te\ and local comparison samples both have mean and median reddening-corrected $\log$\Neiii/\Oiii~$=-1.11$ with 0.05 dex scatter (not corrected for measurement uncertainty). The difference in sample mean is only $0.006\pm0.011$ dex. The \cite{Zeimann2015} results therefore may indicate very rapid evolution from $z\simeq1\rightarrow2$, possibly from differences in nucleosynthetic yields, although this is difficult to reconcile with the lack of evolution at $z<1$. Further measurements at $z>2$ are needed to confirm whether neon emission is indeed enhanced and to determine the physical cause.

\subsection{No evolution in strong-line abundance diagnostics}\label{sec:evolution}

We now examine the quantitative relation between gas-phase metallicity and strong emission line ratios as a function of redshift. \Oiii/\Oii\ and \Neiii/\Oii\ ratios are most easily compared since they vary monotonically with metallicity. We calculate expected metallicity for each galaxy in the \Te\ sample based on measured line ratios and the best-fit relations derived from local galaxies (Table~\ref{tab:diagnostics}), and compare the expected values with those derived in Section~\ref{sec:properties}. We note that measurement uncertainty (particularly in \Oiii$\lambda$4363) results in a skewed metallicity distribution, such that the sample mean is biased toward higher values while the weighted mean is biased toward lower values. The median is robust, however, and we therefore use sample medians for comparison. The median offset at $z\simeq0.8$ is $\Delta \log{\mathrm{O/H}} = 0.01\pm0.03$ for \Neiii/\Oii\ and $\Delta \log{\mathrm{O/H}} = -0.02\pm0.03$ for \Oiii/\Oii. Other line ratios show similarly null evolution of typically $-0.01\pm0.03$ dex. Averaging these measurements does not decrease the uncertainty, which is statistically limited by the sample size. We conclude that there is no evidence for evolution in metallicity at fixed line ratios from $z=0 \rightarrow 0.8$. Local calibrations of the strong line abundance diagnostics considered here therefore appear to be valid to at least $z\simeq1$, within the measurement precision of 0.03 dex in $\log{\mathrm{O/H}}$.

\subsection{No evolution in $\alpha$-element emission line ratios}

\begin{figure*}
\hspace{0.3in}
\centerline{
\includegraphics[width=0.85\columnwidth]{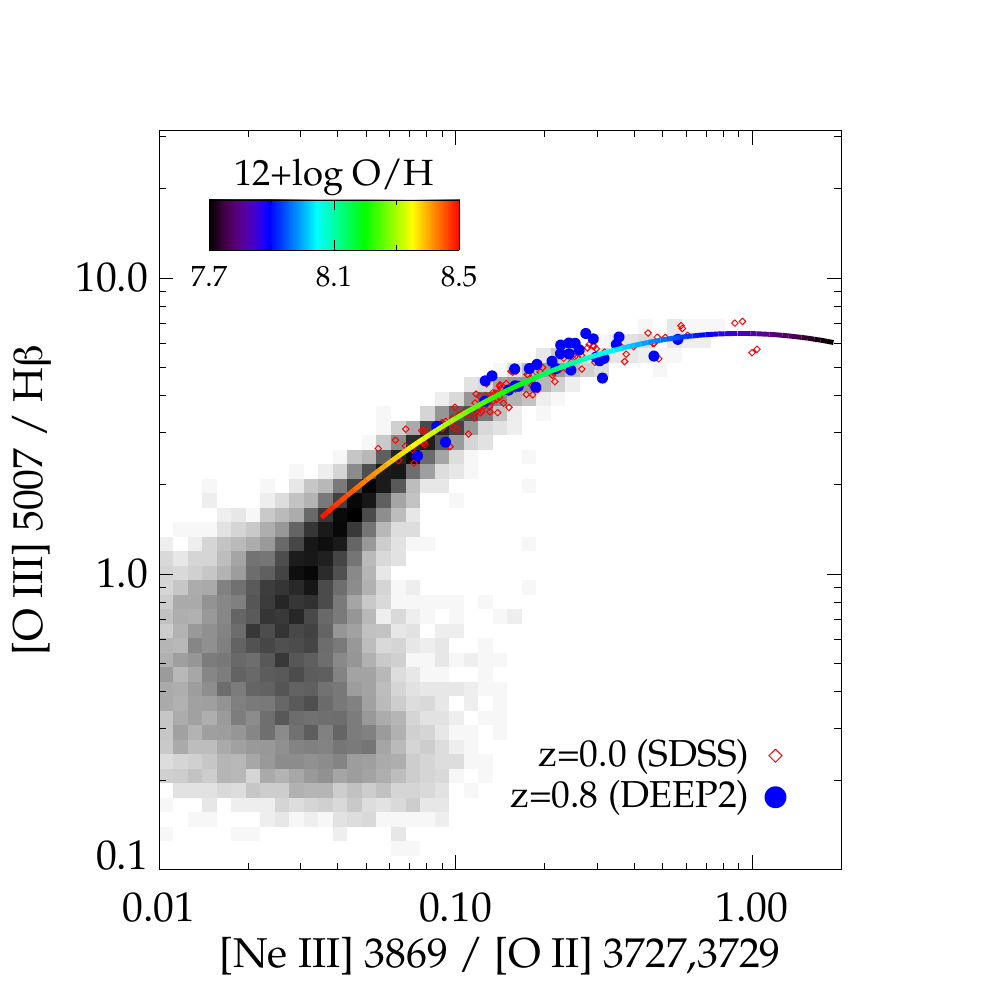}
\hspace{-0.5in}
\includegraphics[width=0.85\columnwidth]{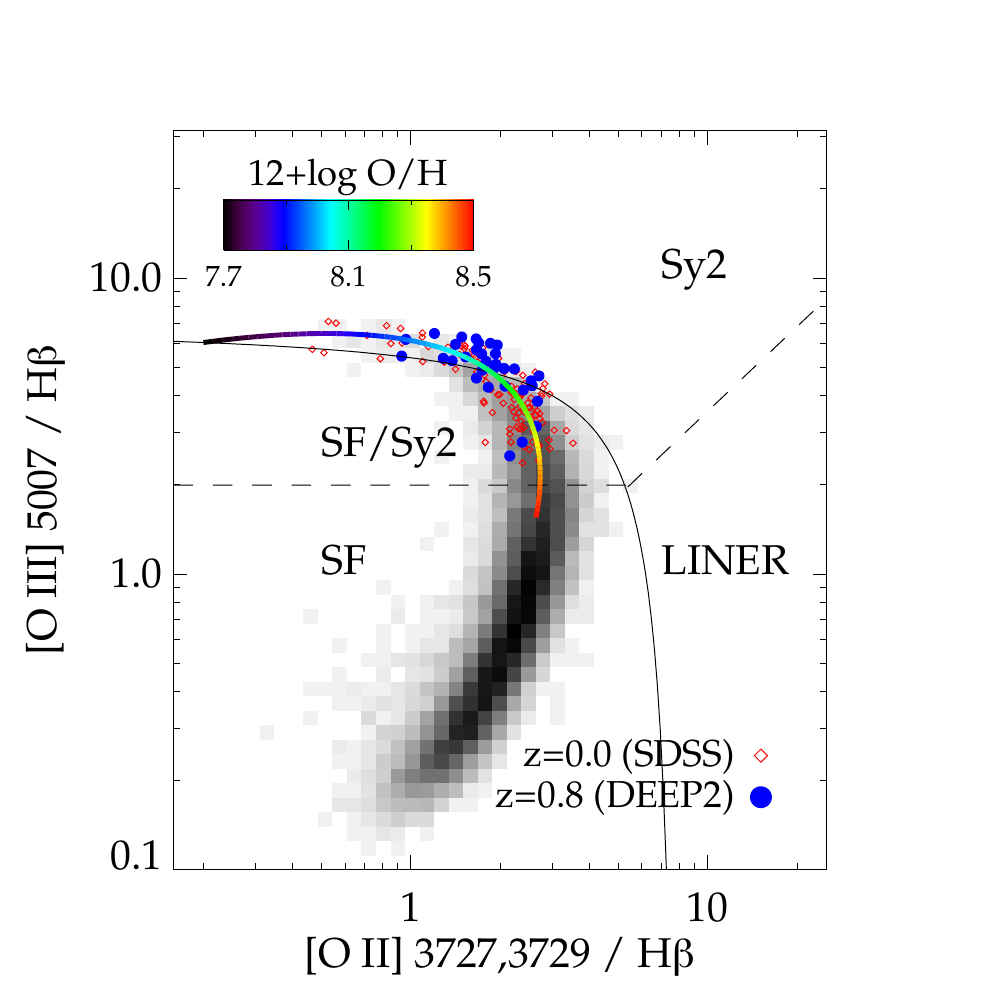}
\hspace{-0.7in}
\includegraphics[width=0.85\columnwidth]{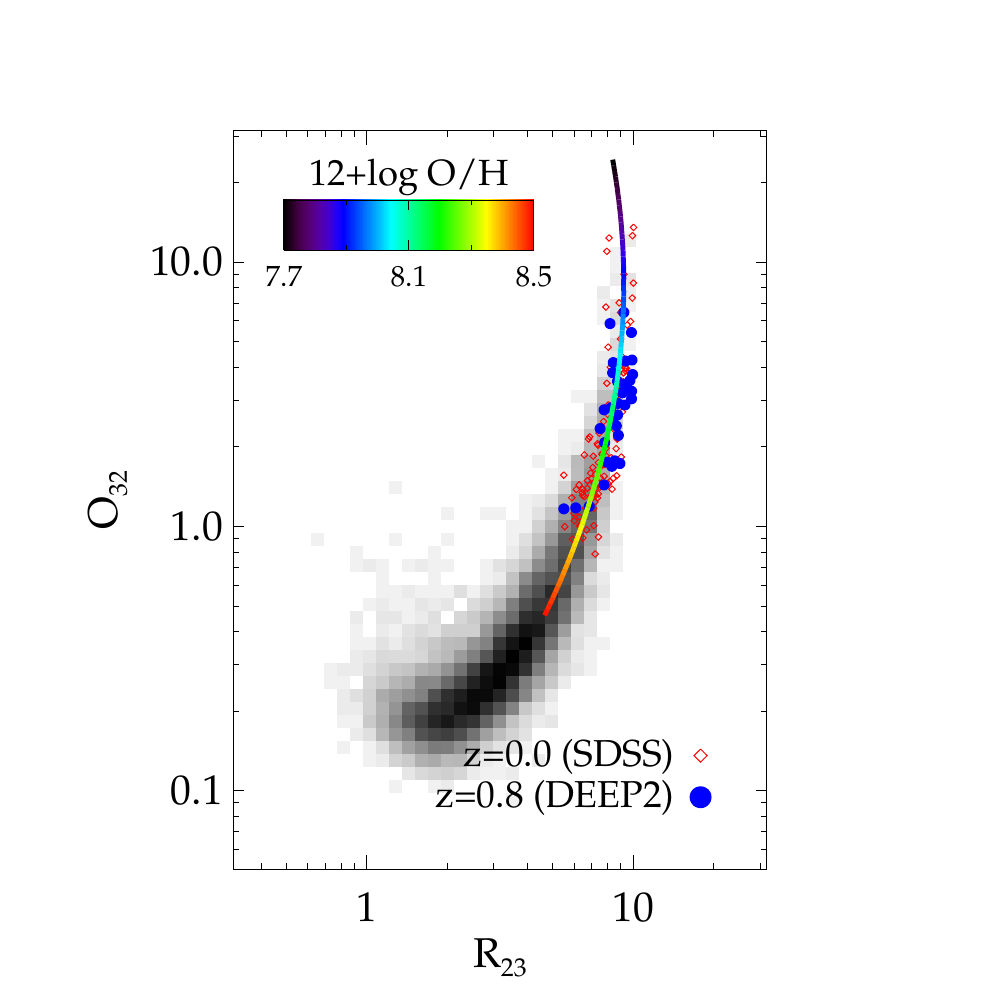}}
\caption{
\label{fig:bpt}
Strong emission line ratios and best-fit metallicity relations (from Table~\ref{tab:diagnostics}). All emission line ratios are corrected for nebular reddening although this correction is negligible for the left-most panel. Galaxies in the \Te\ sample at $z\simeq0.8$ (blue points), the local comparison sample (red points), and the broader population of star forming galaxies in SDSS (grey shading) all populate the same locus with low scatter ($\simeq0.04$ dex RMS). The \Te\ sample is within $0.01\pm0.01$ dex of the locus defined by local galaxies indicating no significant evolution at $z\lesssim1$. The central panel is the ``blue" diagnostic diagram with boundaries described by \cite{Lamareille2010}. We caution that the star forming locus extends into the Seyfert 2 region at low metallicities (\oh~$\lesssim 8.0$).
}
\end{figure*}

We show several line ratio diagnostic diagrams in Figure~\ref{fig:bpt}. Qualitatively it is clear that galaxies in the \Te\ sample at $z\simeq0.8$ follow the same locus as local star-forming galaxies (as is also true for the parent DEEP2 sample, shown in Figure~\ref{fig:o3bias}). This holds for de-reddened flux ratios of emission lines which are widely separated in wavelength (e.g., R$_{23}$ and O$_{32} =$~\Oiii/\Oii) as well as those for which reddening corrections are negligible (\Oiii/\Hb\ and \Neiii/\Oii). Essentially, Figure~\ref{fig:bpt} demonstrates that star-forming galaxies form a tight locus in the parameter space defined by their intrinsic (de-reddened) \Oii, \Oiii, \Neiii, and Balmer emission line fluxes, and that this locus does not evolve with redshift at $z\lesssim1$. As a quantitative example, the \Oiii/\Hb\ versus \Neiii/\Oii\ locus has a dispersion of only 0.04 dex in the \Te\ sample (and 0.03 dex in the local comparison sample), with an offset of $0.01\pm0.01$ dex in \Oiii/\Hb\ at $z\simeq0.08$.

The null evolution in line ratios is not surprising given the lack of any apparent evolution discussed previously in this section. However it is worth emphasizing given that very strong evolution is observed in the parameter space of \Nii, \Oiii, and Balmer lines \citep[with $\sim$0.2--0.4 dex offsets at $z\simeq2.3$;][]{Steidel2014,Shapley2015,Kewley2013}. These same galaxies show no evolution from the locus defined by \Oii, \Oiii, and Balmer lines confirming that the non-evolution seen in Figure~\ref{fig:bpt} holds to at least $z\gtrsim2$. Evolution in \Nii/\Ha\ ratios at high redshift is therefore likely due to systematically higher N/O ratios at high redshift \citep{Shapley2015,Masters2014}, arising from the complicated nucleosynthetic production of nitrogen. In contrast, $\alpha$-capture elements such as oxygen and neon originate almost entirely from core-collapse supernovae and their production is thought to accurately trace the integrated star formation history with little scatter.

Several authors have discussed how the average \Oiii/\Hb\ ratio of star forming galaxies increases with redshift. Possible explanations for this behavior include sample selection bias and evolution in various physical properties of \Hii\ regions (such as ionization parameter, density, and metallicity; e.g., \citealt{Juneau2014,Kewley2013}). While all of these effects are likely present, our results suggest a simple scenario in which galaxies at all redshifts populate a {\em constant locus} of $\alpha$-element and hydrogen line ratios, with evolution in the density of galaxies along the locus (i.e., toward higher \Oiii/\Hb\ ratios at higher redshifts). The locus position is governed largely by metallicity as shown in Figure~\ref{fig:bpt}. (Other properties such as ionization parameter are of course correlated with metallicity.) Therefore, evolution of the galaxy population simply reflects the overall increase in metallicity (O/H) over cosmic time, which has been quantified in numerous studies of evolution in the mass-metallicity relation \citep[e.g.,][and many others]{Maiolino2008,Zahid2013}. These studies indicate an evolution of $\sim$0.1--0.2 dex in metallicity at fixed stellar mass from $z\simeq0\rightarrow0.8$. This amounts to a change in \Oiii/\Hb\ ratios by $\gtrsim$0.2 dex for typical galaxies. We note that in contrast to the diagrams in Figure~\ref{fig:bpt}, loci of \Oiii/\Hb\ as a function of \Nii/\Ha\ or stellar mass \citep[the Mass-Excitation diagram;][]{Juneau2014} are {\em not} constant because of evolution in N/O abundances and stellar mass at fixed O/H.

\section{Discussion}\label{sec:discussion}

\subsection{Metallicity inferences at high redshift}

The primary purpose of this paper is to present a set of strong-line metallicity diagnostics that can be accurately used at high redshifts. Ultimately the diagnostics given in Table~\ref{tab:diagnostics} and Figure~\ref{fig:offset} are calibrated from local data since we find that data at $z\simeq0.8$ give fully consistent results but with larger uncertainty. While this study is limited to $z\lesssim0.8$, galaxies at $z\simeq2.3$ show no offset in the line ratios considered here (\citealt{Shapley2015}; but see also \citealt{Zeimann2015} regarding \Neiii). We therefore consider it likely that these diagnostics are accurate up to $z>2$ and perhaps even at all redshifts, although this has yet to be confirmed with direct \Te-based measurements. We wish to emphasize that the diagnostics in Table~\ref{tab:diagnostics} are only valid for the range of line ratios and metallicities probed by this work (\oh~$=7.8$--8.4). Figure~\ref{fig:bpt} suggests that mild extrapolation may yield reasonable results but we do not endorse this.

A broader consequence of the non-evolution in strong-line metallicity diagnostics (those in Table~\ref{tab:diagnostics} and Figure~\ref{fig:offset} at least) is that {\em any} local calibration can be used to infer self-consistent metallicity evolution from high redshift data. A variety of such calibrations exist in the literature based on different methods including the ``direct" \Te\ method, photoionization models, and recombination lines. Different methods are well known to give discrepant results \citep[e.g.,][]{Blanc2015} and hence the true absolute metallicity scale remains under debate. Even the calibrations presented here are subject to systematic variation by up to 0.05 dex in metallicity depending on the adopted atomic parameters and \Te(\Oii)--\Te(\Oiii) relation. However, the \Te-based analysis in this work demonstrates clearly that metallicity does not evolve at fixed strong line ratios, at least for the hydrogen and $\alpha$-element lines considered here.

We advocate using multiple diagnostics simultaneously to infer metallicity from strong-line methods. It is clear from Figures~\ref{fig:offset} and \ref{fig:bpt} that the usefulness of various diagnostics is a strong function of their position along the locus of line ratios (i.e., metallicity). For example, O$_{32}$ gives a more precise constraint on metallicity than R$_{23}$ in the regime where O$_{32} > 1$ (\oh~$<8.4$), while R$_{23}$ is the more precise indicator at higher metallicities. In any case the precision is clearly improved when using both in combination (right-most panel of Figure~\ref{fig:bpt}). The best approach is of course to use {\em all} available diagnostics to infer metallicity, as was done for example in \citet{Belli2013} and \citet[][with an excellent illustrative example in Figure 6 of that paper]{Maiolino2008}. One may also use observed (reddened) flux ratios combined with priors such as \Hg/\Hb~$=0.47$ to produce a 2-dimensional probability distribution in both reddening and metallicity \citep[e.g.,][]{Maiolino2008}, which is more robust than adopting a fixed reddening estimate. We note that while \Neiii/\Oiii\ is insensitive to metallicity, it provides a valuable constraint and sanity check on the nebular reddening (Figure~\ref{fig:rv}).
Knowledge of the intrinsic scatter is essential for combining multiple diagnostics: the scatter allows for a straightforward calculation of the likelihood of any metallicity and nebular reddening (i.e., the posterior probability distribution) from a given set of emission line ratios, from which the most likely metallicity and confidence intervals can be computed. It is for this reason that we list the intrinsic scatter in line ratios at fixed metallicity for each of our calibrations in Table~\ref{tab:diagnostics}.

It is worth considering whether the precision of strong-line metallicity methods can be improved by introducing a second parameter. For example, photoionization models suggest that the intrinsic scatter in metallicity calibrations can be reduced by fitting for the ionization parameter in addition to metallicity \citep[e.g.,][]{Pilyugin2001,Kobulnicky2004,Blanc2015}. Typically the idea is to use e.g. R$_{23}$ primarily as an indicator of metallicity, with a correction for ionization parameter derived from additional line ratios such as O$_{32}$. We conducted an empirical test by measuring the correlation between metallicity and offsets in the direction orthogonal to the best-fit loci shown in Figure~\ref{fig:bpt}. There is a significant trend in the local comparison sample, in the sense that
\begin{equation}\label{eq:offsets}
\frac{\Delta\log{\mathrm{O/H}}}{\Delta\log{\mathrm{R_{23}}}} = 1.89\pm0.14
\end{equation}
at constant O$_{32}$. The error represents formal uncertainty in the best-fit slope and does not reflect the much larger sample variance. While this indicates a statistically significant secondary parameter, its effect is negligible: applying Equation~\ref{eq:offsets} reduces the intrinsic scatter in inferred metallicity by only 3\% compared to using the O$_{32}$ calibration alone. 
Likewise the $f(\mathrm{P,R_{23}})$ method of \cite{Pilyugin2001} does not substantially reduce the scatter.
We therefore find no compelling reason at this time to include secondary trends between metallicity and offsets from the line ratio loci in Figure~\ref{fig:bpt} when calculating metallicity using the strong-line calibrations presented here.

\subsection{Prospects for \Te-based metallicities at $z>2$}

As the first such work at cosmological distances, our investigation of strong-line metallicity diagnostics at $z\simeq0.8$ provides helpful guidance for future efforts to undertake a similar study at higher redshifts. It is imperative to construct a sample which is complete in sensitivity, without regard to emission line detection significance as this induces a bias. 
We note that \Te-based metallicities for $z\gtrsim1.5$ galaxies reported in the literature \citep[by][]{Villar2004,Yuan2009,Christensen2012,James2014,Steidel2014} are on average lower than expected from their strong line ratios, although most individual measurements are compatible with the $z=0$ relations given the scatter. However the existing sample at these redshifts is too small for robust conclusions.
For our sample, selecting only the galaxies with $\geq3\sigma$ detections of \Oiii$\lambda$4363 in Figure~\ref{fig:sample} would lead us to conclude that metallicity calibrations evolve by $\Delta\log{\mathrm{O/H}} \simeq -0.2$ dex at fixed \Oiii/\Oii\ compared to $z=0$. Even requiring a 5$\sigma$ detection would lead to an inferred evolution of $\gtrsim$0.1 dex. A sensitivity such that uncertainty in the ratio of \Oiii$\lambda$4363/\Oiii$\lambda5007$ is $\sigma\lesssim0.003$ is sufficient for a sample with comparable (or lower) metallicity to that used here. Our sample of 32 galaxies yields a precision of 0.03 dex in the evolution of strong-line metallicity diagnostics, and hence even a modest sample of 10-15 galaxies would provide a measurement of evolution at the $\sim$0.05 dex level. Given the wealth of spectroscopic data now in hand from ground- and space-based surveys \citep[e.g.,][]{Steidel2014,Shapley2015,Maseda2014}, it may be practical to construct such a sample at $z\gtrsim2$ with careful pre-selection and moderately deep followup to obtain the requisite sensitivity to \Oiii$\lambda$4363.

\subsection{Implications for evolution in the BPT diagram}

Our choice of \Oiii/\Hb\ as the abscissa in Figure~\ref{fig:nt} is motivated by the strong observed redshift evolution in either \Oiii/\Hb\ or/and \Nii/\Ha\ (the classic BPT diagram). We have found no evidence for changing physical conditions (specifically \Ne, \Te, O/H, and Ne/O) in galaxies with fixed \Oiii/\Hb\ flux ratios. Furthermore we find no offset from the locus of line ratio diagrams involving \Oiii/\Hb\ and other $\alpha$ elements, with $\simeq$0.01 dex precision (Figure~\ref{fig:bpt}). We therefore strongly suspect that evolution in the BPT diagram is due to \Nii/\Ha\ and not \Oiii/\Hb. A straightforward explanation is that N/O abundance ratios are systematically higher in high redshift galaxies at fixed $\alpha$-element line ratios (i.e., fixed metallicity, as considered by \citealt{Masters2014,Shapley2015,Steidel2014}). This appears to be a plausible explanation, since relatively nearby galaxies with properties similar to high redshift samples (such as the ``green peas" and ``Lyman break analogs") have higher N/O than expected for their O/H metallicity \citep[e.g.,][]{Izotov2011,Amorin2010}. This can result from various effects related to high star formation rates such as metal-poor gaseous inflows, metal-rich outflows, a large population of Wolf-Rayet stars, and previous star formation history \citep[e.g.,][]{Amorin2010,Masters2014,Andrews2013,Martin2008}.

It is not yet known whether the galaxies in our sample at $z\simeq0.8$ are offset from local galaxies in terms of \Nii/\Ha. Near-IR spectroscopic followup with Keck is ongoing to measure \Nii, \Ha, temperature-sensitive \Oii$\lambda\lambda$7320,7330, and other diagnostic features. Data gathered from the followup campaign will provide accurate \Te-based measurements of N/O and determine whether nitrogen abundance variations can plausibly explain systematic offsets toward higher \Nii/\Ha\ flux ratios as observed in high redshift galaxies.

\section{Conclusions}\label{sec:conclusions}

We have examined the relation between various strong optical emission line ratios and direct \Te-based measurements of metallicity (gas-phase oxygen abundance) in a homogeneous sample of star forming galaxies at $z=0.72$--0.87 drawn from the DEEP2 survey. While these strong-line diagnostics are the principal method used to infer galaxy metallicities beyond the local volume, this work represents the first direct calibration at cosmological distances. Our results confirm that at least some of the commonly used diagnostics are valid up to $z\lesssim1$ and provide guidance for future studies at higher redshifts.

Our main results are as follows:
\begin{itemize}
\item The relation between metallicity and strong emission lines of oxygen, hydrogen, and neon does not evolve between $z=0\rightarrow0.8$, to within our measurement precision of 0.03 dex in O/H ($1\sigma$). The relation between Ne/O and O/H abundance is likewise constant: we measure $\Delta\log{\mathrm{Ne/O}} = 0.01\pm0.01$ at fixed O/H at $z\simeq0.8$ relative to $z=0$. Therefore, these locally calibrated strong-line metallicity diagnostics are valid to at least $z\simeq0.8$, and they may remain accurate even at the highest redshifts.

\item We present a set of calibrations between metallicity and strong line ratios with the intent of using multiple such diagnostics simultaneously. Our calibrations define a locus in the parameter space of \Oii, \Oiii, \Neiii, and Balmer emission lines which is designed to closely follow the observed locus of star forming galaxies. Via these diagnostics, metallicity can be accurately inferred from the position of a galaxy on various two-dimensional line ratio diagrams (e.g., O$_{32}$ vs. R$_{23}$) with improved precision compared to using a single diagnostic such as R$_{23}$. We include measurements of the intrinsic scatter so that multiple diagnostics can be combined to calculate a posterior probability distribution in metallicity (and nebular reddening), given a set of emission line ratios.
\\
We find evidence for a secondary parameter in the sense that metallicity is correlated with a galaxy's offset from the locus. However, including this offset as an additional parameter does not substantially reduce the scatter in our strong-line calibrations.

\item Star forming galaxies at $z\simeq0.8$ follow the same locus of oxygen, hydrogen, and neon emission line ratios as at $z=0$, with offsets of $\sim0.01\pm0.01$ dex and low scatter ($\lesssim$0.04 dex). This is in contrast to the star formation locus of \Nii/\Ha\ versus \Oiii/\Hb, which has larger scatter and evolves with redshift (by $\gtrsim$0.2 dex at $z=2$). Our results as well as earlier work \citep{Shapley2015,Masters2014} suggest that this evolution is exclusively in \Nii/\Ha, possibly caused by N/O abundance variations. Ongoing followup spectroscopy is needed to confirm or refute this hypothesis within our sample.

\end{itemize}

While this work represents an important first step toward confirming the validity of strong-line methods for inferring metallicity in high redshift galaxies, we are acutely aware of the limited range in redshift and metallicity probed by our \Te\ sample. The narrow $z=0.72$--0.87 is restricted solely by the availability of spectra with sufficient sensitivity, spectral resolution, and wavelength coverage. Extending this work to higher redshifts will require suitable spectra at near-infrared wavelengths. Encouragingly, near-IR spectra now exist for large samples of galaxies at $z=1$--3.5 thanks to dedicated surveys with multiplexed instruments such as Keck/MOSFIRE, VLT/KMOS, and HST/WFC3 \citep[e.g.,][]{Steidel2014,Kriek2014,Wisnioski2015,Brammer2012}. These surveys are not deep enough to detect \Te-sensitive emission lines, however, and so additional followup will be necessary for \Te-based studies. It should already be practical to select modest sub-samples for efficient followup using a selection method similar to ours (see Figure~\ref{fig:sample}), requiring strong \Oiii$\lambda\lambda$4959,5007 lines and favorable redshifts for observing \Oiii$\lambda$4363. The work of \cite{Maseda2014} illustrates the potential efficacy of this approach. Pre-selection of \Oiii\ emitters from HST grism spectra offers an additional advantage over purely ground-based spectroscopy by including redshifts at which the strong lines are affected by telluric features (for example $z\simeq1.65$--2.0, where \Oiii$\lambda$4363 is in the favorable J-band window while \Oiii$\lambda$5007 and \Hb\ are obscured by telluric absorption).

It is somewhat fortuitous that our calibrations cover only \oh~$\lesssim8.4$, since \Te-based results and other methods become increasingly discrepant at higher metallicity. There are practical concerns as well: \Oiii$\lambda$4363 emission is progressively weaker at higher metallicity due to a combination of decreasing \Hii\ region temperature and a lower O$^{++}$ fraction. At higher metallicities, oxygen abundance is dominated by the O$^+$ ion {\em and} the temperature-sensitive \Oii$\lambda\lambda$7320,7330 lines become stronger than \Oiii$\lambda$4363 \citep[at \oh~$\gtrsim8.2$--8.4; e.g.,][]{Izotov2006,Andrews2013}. Therefore, selecting galaxies with strong \Oii\ emission and precise measurements of \Te(\Oii) may provide a practical way to extend our study to higher metallicities. 
Additionally the \Te-sensitive sulfur lines \Sii$\lambda\lambda$4069,4076 and \Siii$\lambda$6312 may be useful \citep[e.g.,][]{Peimbert1969,Perez-Montero2003}.
Stacked spectra are another possible way forward \citep[e.g.,][]{Andrews2013}.

Throughout this work we have been careful to compare data at $z\simeq0$ and $z\simeq0.8$ using consistent methods. Our results regarding redshift evolution are therefore robust. However, we caution that our \Te-based methods do not necessarily reflect the absolute metallicity scale. Different assumptions for ionization corrections, the \Te(\Oii)-\Te(\Oiii) relation, and atomic properties can plausibly change the derived metallicities by $\sim$0.05 dex. Different methods (e.g., based on photoionization modeling or recombination lines) suggest even larger differences of $\sim$0.2 dex. Discrepancies between these methods can plausibly be reconciled by spatial variations in temperature or by a non-Maxwellian electron energy distribution \citep[e.g.,][]{Dopita2013}. If the cause can be identified and characterized, then optimistically the \Te\ method and associated strong-line diagnostic calibrations can be updated to produce truly reliable metallicity measurements. For now, we can at least rule out any strong systematic evolution with redshift for the strong-line diagnostics considered here.

\acknowledgements

We thank the referee for several helpful suggestions. 
We gratefully acknowledge Renbin Yan and Jeff Newman for their assistance with spectroscopic throughput corrections. 
TAJ acknowledges support from the Southern California Center for Galaxy Evolution through a CGE Fellowship. 
CLM acknowledges support from the National Science Foundation under AST-1109288.
Funding for the DEEP2 Galaxy Redshift Survey has been provided by NSF grants AST-95-09298, AST-0071048, AST-0507428, and AST-0507483 as well as NASA LTSA grant NNG04GC89G. 
Spectroscopic data in DEEP2 were obtained with the DEIMOS spectrograph at the W. M. Keck Observatory, which is operated as a scientific partnership among the California Institute of Technology, the University of California, and the National Aeornautics and Space Administration. The Observatory was made possible by the generous financial support of the W. M. Keck Foundation. 

\end{document}